\documentclass{article}
\usepackage{authblk}
\usepackage{fullpage}

\usepackage{amssymb}
\usepackage{graphicx} 
\usepackage{xcolor}
\usepackage{url} 
\usepackage{booktabs}
\usepackage[linesnumbered,ruled,vlined]{algorithm2e}
\usepackage{pgfplots}
\usepackage{tikz}
\usetikzlibrary{shapes.geometric}
\usetikzlibrary{arrows.meta}
\usepackage{float}
\usepackage{subcaption}
\usepackage{comment}
\usepackage{yhmath}
\usepackage{ifthen}
\usepackage{microtype}

\newboolean{useInline}
\setboolean{useInline}{true} 

\ifthenelse{\boolean{useInline}}
  {\usepackage[appendix=inline]{apxproof}}
  {\usepackage[appendix=append]{apxproof}}
\usepackage{hyperref}

\newcommand{\LCM}{{\em Look-Compute-Move}}
\newcommand{\seq}{$\mathcal{SQ}$}
\newcommand{\fsync}{$\mathcal{FSYNC}$}
\newcommand{\ssy}{$\mathcal{SSYNC}$}
\newcommand{\async}{$\mathcal{ASYNC}$}
\newcommand{\R}{\mathcal{R}}

\newcommand{\map}{\Gamma}
\newcommand{\cir}{Cir}
\newcommand{\Cir}{\mathcal{r}}
\newcommand{\Uni}{Universal}
\newcommand{\UPF}{{\tt UPF}}
\newcommand{\WPF}{{\tt WPF}}

\newcommand{\OB}{$\mathcal{O}${\sc BLOT}}

\newcommand{\sepa}{\sc Separate}
\newcommand{\lead}{\sc Leader}
\newcommand{\overl}{\sc Overlap}
\newcommand{\occ}{\sc Occupy}
\newcommand{\last}{\sc Last}
\newcommand{\spf}{\sc SqPF}
\newcommand{\smpf}{\sc SqPF$'$}

\newtheoremrep{theorem}{Theorem}
\newtheoremrep{lemma}{Lemma}
\newtheoremrep{corollary}{Corollary}
\newtheoremrep{property}{Property}

\title{Universal Pattern Formation by Oblivious Robots Under Sequential Schedulers\thanks{Some of the results contained in this paper were presented at the 32nd International Colloquium On Structural Information and Communication Complexity (SIROCCO)~\cite{FNPPS2025}. 
This work has been  funded in part   
by the Italian National Group for Scientific Computation INdAM - GNCS Project, CUP E53C24001950001, and by the Natural Sciences and Engineering Research Council of Canada, under Discovery Grants A2415 and 203254.
}
}

\author[1]{Paola Flocchini}
\author[2]{Alfredo Navarra}
\author[3]{Debasish Pattanayak}
\author[2]{Francesco Piselli}
\author[4]{Nicola Santoro}

\affil[1]{School of Electrical Engineering and Computer Science, University of Ottawa, Ottawa, Canada\\ \texttt{paola.flocchini@uottawa.ca}}
\affil[2]{Department of Mathematics and Computer Science, University of Perugia, Perugia, Italy\\ \texttt{alfredo.navarra@unipg.it, francesco.piselli@unifi.it}}
\affil[3]{Department of Computer Science and Engineering, Indian Institute of Technology Indore, Indore, India\\ \texttt{debasish@iiti.ac.in}}
\affil[4]{School of Computer Science, Carleton University, Ottawa, Canada\\ \texttt{santoro@scs.carleton.ca}}

\date{}

\begin{document}
\title{Universal Pattern Formation by Oblivious Robots Under Sequential Schedulers
}


\maketitle

\abstract{
 We study the computational power that oblivious robots operating in the plane have under \emph{sequential} schedulers. We show that this power is much stronger than the obvious capacity these schedulers offer of breaking symmetry, and thus to create a leader. In fact, we prove that under any sequential scheduler, robots are capable of solving problems that are unsolvable even with a leader under the fully synchronous scheduler $\mathcal{FSYNC}$.
More precisely, we consider the class of \emph{pattern formation} problems, and focus on the most general problem in this class, {\tt Universal Pattern Formation} ({\tt UPF}), which requires the robots  to form \emph{every} pattern given in input, starting from \emph{any} initial configuration (where some robots may occupy the same point, hence forming a \emph{multiplicity}).

We first show that {\tt UPF} is \emph{unsolvable} under  $\mathcal{FSYNC}$, even if the robots are endowed with additional strong capabilities (multiplicity detection, rigid movement, agreement on coordinate systems, presence of a unique leader). On the other hand, we prove that, except for point formation ({\tt Gathering}),  {\tt UPF}  is \emph{solvable} under any sequential scheduler without any additional assumptions.
We then turn our attention to the {\tt Gathering} problem, and prove that weak multiplicity detection (the ability to detect a multiplicity but not the exact number of robots forming it) is necessary and sufficient for solvability under sequential schedulers. The results obtained show that the computational power of the robots under $\mathcal{FSYNC}$ (where {\tt Gathering} is solvable without any multiplicity detection) and that under sequential schedulers are \emph{orthogonal}.}

\section{Introduction}

Distributed systems composed of a set $\mathcal{R}$ of autonomous mobile robots operating in Euclidean spaces  have been the object of rather extensive investigations in theoretical computer science,
since their introduction in  \cite{SuzukY99}.

The robots operate performing \LCM\  cycles; in each cycle, a robot  observes the location of the other robots ({\em Look}), executes its algorithm to determine a destination ({\em Compute}), and moves toward the computed destination ({\em Move}).  
Several formal models have been proposed, describing different levels of computation capabilities of the robots. In this paper, 
we consider the
standard model of silent oblivious robots, $\mathcal{OBLOT}$, which describes  computationally weak robots (e.g., see \cite{CanepDIP16,FloccPS12,IzumiSKIDWY12,YamasS10}), and our focus is on deterministic computations.

As for other distributed systems,   the main research objective
is  the determination of what is computable, under what conditions, and at what cost; the main goal is to identify the minimal capabilities needed by the robots for the system to  be able to solve a given problem or perform an assigned task.
 The feasibility of a task and the solvability of a problem are influenced by several specific  factors;
the most determinant one 
concerns
 the nature of the
timing of the  robots actions  imposed  by the adversarial  environment in which the robots operate.  

\subsection{Sequential Adversarial Schedulers}
%
The timing of when a robot is active and performs a cycle as well as the duration of each action  within a cycle is determined by an adversary, called 
{\em scheduler}. There are two general classes of schedulers, 
the asynchronous and the synchronous ones; in both classes, all schedulers
must respect the {\em fairness} constraint: every robot is activated 
within finite time, 
infinitely often.

In the {\em asynchronous} class ({\sc Async}), each robot may be activated at arbitrary times, independently of the others, and the duration of each operation in the cycle is finite but possibly unpredictable, and might be different in different cycles.  In this class, the standard scheduler investigated in the literature is the one 
that does not have any additional restrictions besides fairness (hence said to have unrestricted adversarial power); 
called simply  ({\em fully}) {\em  asynchronous} and  denoted by \async, it was introduced in \cite{FloccPSW99}. 
A few other schedulers in this class have also been investigated
 (see, e.g., \cite{CicerDN21,FloccSSW23,KirkpKNPS24}).

In the {\em synchronous} class, ({\sc Synch}  $\subset$ {\sc Asynch}),
time is logically divided into a sequence of \emph{rounds}. 
In each round, the scheduler activates a non-empty subset of robots, and they perform each phase of their  \LCM\ cycle simultaneously, in perfect synchrony. 
Thus, the only distinction between  synchronous schedulers, is their {\em adversarial power}, i.e., the amount of freedom  each  has in the choice of which robots to activate in each round.
In this class, the standard scheduler is  the one with unrestricted adversarial power (i.e., the one 
that does not have any additional restrictions  besides synchrony and fairness);  known as  
{\em semi-synchronous} and denoted by \ssy, it has been introduced in \cite{SuzukY99}.
At the other end of the spectrum, there is the  very popular {\em fully-synchronous} scheduler, denoted by \fsync,
where all robots are activated in every round.

Observe that  \fsync\  has no adversarial power, 
 with respect to the activation of the robots.
An apparently intuitive consequence of this observation is  that, thus,  
the computational power of the robots under \fsync\ is  the strongest\footnote{I.e., the problems solvable under \fsync\  include all the ones solvable under any other synchronous scheduler.} possible. Such an intuition is however erroneous.
Indeed there are  schedulers in {\sc Synch} that allow robots to solve important problems unsolvable under \fsync.
Such is the subclass of {\em sequential} schedulers, {\sc Sequential} $\subset$ {\sc Synch}, where only a single robot is activated in each round. 

An immediate obvious computational  advantage that the robots have under such schedulers  is that the robots  can then break symmetry (and thus, e.g., elect a leader), a task impossible to achieve under \fsync\ when starting from a symmetric configuration. 

 In spite of this remarkable fact, the actual computational relationship between \fsync\ and the class of sequential schedulers has so far been ignored. In general, very little  is known on the computational power of robots under sequential schedulers. This is in contrast with the interest shown on sequential schedulers in other fields, e.g., in  self-stabilization (where they are called {\em centralized}) and in the theory of cellular automata (where they are called {\em asynchronous} \cite{Fa2018}). In the field of mobile robots  they have been analyzed only in the area of fault-tolerance   
limited to   the problems of gathering and scattering  (see \cite{DefagPT19} for a recent survey),  
and gathering  under a  round robin scheduler   \cite{FreiW24,NP25}.

 Motivated by the  existing knowledge gaps on the computational power of robots operating under sequential schedulers, in this paper, we investigate the large and important class  of {\em Pattern Formation} problems to determine the  impact  that these schedulers have on the resolution of such problems. In particular, we focus on the ability of robots to solve those problems under the {\sc Sequential} scheduler with the strongest adversarial power, henceforth denoted by {\seq}, and contrast it with that of \fsync.

\subsection{Pattern Formation}
%
{\em Pattern formation}  refers to  the abstract problem of the robots rearranging themselves so that, at some point in time, the set of their locations   
in the plane (the {\em configuration})
satisfies some  geometric predicate given in input (the {\em pattern}) and no longer move. Typically, the pattern is specified as a set of points $\hat{P}$  representing some geometric figure
that the robots must form (irrespective of rotation, reflection, translation or scaling). 
In some applications, forming a specific pattern  (e.g., a line, a circle, as in~\cite{DefagS08,DieudLP08,FloccPSV17}) may be the first step of a more complex task  (e.g., flocking, sweep, search, etc.).

Due to their importance, the class of pattern formation problems has been extensively studied under a number of different assumptions.
A very special  role in this class is  held by  {\em point formation};  that is, the  problem of forming  the pattern consisting of a single point, i.e., $|\hat{P}| =1$.  
It is relevant  because it corresponds to the important coordination problem, extensively studied in the literature,
called  {\tt Gathering} or {\tt Rendezvous}, requiring all  robots to meet at the same location, not fixed in advance (see, e.g.,~\cite{AgmonP06,CieliFPS12,Flocc19,KameiLOT11,PattaAM21,PattaMHM19,Prenc07}). 
The class of pattern formation problems is quite large, not only because of the different types of geometric patterns that can be formed under different conditions, many requiring special ad-hoc techniques for their formation  (e.g., see \cite{CicerDN19,DefagS08,DieudLP08,FloccPSV17}),  but also for the wide range of more general research questions it poses. These include, for  a specific adversarial scheduler:  under what conditions can all patterns be formed, what  patterns are formable from a given initial configuration,  what patterns are formable from any initial configuration. In such investigations,  the  positive answers  usually  
 depend on the symmetry of the pattern to be formed and that of the initial positions of the robots,
and are established through the  design of  algorithms that allow the robots to form the claimed patterns.

Clearly, to form a pattern, there must be at least as many  robots as there are points in the pattern, i.e.,  $|{\cal R}| \geq |\hat{P}|$; we shall call this the {\em trivial assumption}.
Most of the research has however relied on a variety of  non-trivial assumptions, such as: 
 the robots occupy distinct initial locations;   the number of robots  is equal to the number of points in the pattern, $|{\cal R}| =  
 |\hat{P}|$; the robots have some agreement on coordinate systems or chirality; 
 one of the robots is visibly different from all the others (i.e., there is a {\em leader}), etc. (see, e.g.,~\cite{FloccPSV17,CicerDN19a,CicerDN19,CicerSN21,DPV10,FloccPSW08}).

In this paper we are interested in the most  general problem in this class,  {\tt \Uni\  Pattern Formation}  ({\tt UPF}), which requires
the robots  to form {\em every} pattern given in input  starting from {\em any} initial configurations, regardless of their number, of the locations they initially occupy, 
and of the number of points in the pattern, and we are interested in satisfying only the trivial assumption with the minimal robots capabilities.

\subsection{Contributions}
\label{sec:contributions}
%
In this paper we bring to light the  very strong  computational  power that robots have  under the {\sc Sequential} class  of  schedulers in regards to pattern formation problems, and analyze the adversarial power of  sequential schedulers  in comparison to that of the  other synchronous schedulers studied in the literature.

We do so by focusing on the  general 
{\tt UPF} problem in the case where $|\hat{P}|>1$, i.e., the point formation is not included.
We prove,  in Section~\ref{sec:impossibility}, that this problem is {\em unsolvable}  under \fsync\ (and, thus, under \async\ and \ssy), even if the robots are empowered with strong multiplicity detection (i.e., robots can detect the exact number of robots occupying the same location), their movements are rigid (i.e., they always reach their destination regardless of the distance),  have total agreement on the coordinate system, and there is a unique leader robot (i.e., distinguishable  from all others).

On the other hand, in Sections \ref{sec:P4} and \ref{sec:P5} we prove that, again except for point formation, the {\tt \Uni\  Pattern Formation} problem is {\em solvable}  under any sequential schedulers, regardless of any symmetry of the pattern or of the initial configuration of the robots, and without any additional assumptions on the computational capabilities of the robots, namely even if they are incapable of multiplicity detection, their movements are non-rigid, they have no agreement on the coordinate system nor on chirality, and are all identical. 
The proof is constructive: we present a solution protocol, prove its correctness,  and analyze its time complexity.
These results indicate the strong impact that the presence of a sequential scheduler can have on the 
computational power of the robots, allowing them to solve, with minimal capabilities, problems unsolvable even under \fsync\  with stronger power and a leader. 

Then, in Section~\ref{sec:gathering}, we turn our attention to the point formation or {\tt Gathering} problem. This is known to be {unsolvable} 
under \ssy\ even with strong multiplicity detection (i.e., a robot can detect the exact number of robots in each point). In contrast, we prove that, under any sequential scheduler, the  problem   becomes  solvable with weak multiplicity detection (i.e., a robot can detect only whether a point is occupied by one or more than one robot).
We also prove that, without   weak multiplicity detection, the problem, while solvable   under  \fsync,  is unsolvable under  \seq. 
The results obtained prove  that  the computational power of the robots under \seq\ and 
that under  \fsync\  are {\em orthogonal}.

\section{Model and Terminology}\label{sec:model}
We consider the standard  ${\mathcal {OBLOT}}$ model of 
distributed systems of  autonomous mobile  robots (e.g.,  see \cite{FloccPS12}). 

\subsection{Robots}

The system is composed of a set  
 $\R = \{r_1, r_2, \dots, r_n\}$ of
  computational {\em robots}, 
   moving and operating in the Euclidean plane $\mathbb R^2$.
   Viewed as  points, the robots can move freely and continuously on the plane.
A robot is endowed with motorial and computational abilities, a private coordinate system of which it perceives itself at the origin, and sensorial devices that allow it to observe the positions of the other robots with respect to its local coordinate system.
When active, a robot  performs a \emph{Look-Compute-Move} cycle: 
\begin{enumerate}
\item {\em Look:} The robot  obtains an instantaneous snapshot of the positions occupied by the robots 
expressed within a private coordinate
system;
\item {\em Compute:} The robot executes its built-in deterministic algorithm, the same for all robots,  using the snapshot as input; the result of the computation is a destination point;
\item {\em Move:} The robot moves toward  the computed destination.
If the destination is the current location, the robot stays still.
\end{enumerate}
The duration of each cycle is  finite.
Robots are  {\em  autonomous}:  they operate  without a central control or external supervision;  {\em identical}:
  they are  indistinguishable by their appearance 
and  do not have  
distinct identities that can be used during the computation;
{\em disoriented}: they may not agree on a common coordinate system nor on  chirality (a clockwise orientation of the plane);
 {\em silent}: they have no means of direct
communication of information to other robots,
so any communication occurs in a totally implicit manner, by
moving and by observing the  positions of the other robots;   and they are {\em oblivious}:
at the beginning of each cycle, 
a robot has no memory of past actions and computations, 
so the computation is based solely on what determined in the current cycle. 

Movements are said to be {\em rigid} if the robots always reach their destination in the same round they started moving.
They are said to be {\em non-rigid}  if they may be unpredictably stopped by an adversary 
whose only limitation is the existence of $\delta>0$, unknown to the robots, such that
  if the destination is at distance at most $\delta$, the robot will  
reach it, otherwise it will move at least $\delta$ toward the
  destination. In this work we consider only non-rigid movements.

More than one robot can be at the same point at the same time; such a point is said to be a \emph{multiplicity}. The robots are said to be capable of {\em multiplicity detection} if they can determine whether or not points are multiplicities;  multiplicity detection is said to be {\em strong} if a robot can determine the exact number of robots located at a point, {\em weak} otherwise. 
 In this work, we consider robots without any type of multiplicity detection.

\subsection{Schedulers}

The timing of when a robot becomes active as well as the duration of {\em Compute} and {\em Move} phases within each cycle, and the rate of motion within each {\em Move} phase, is viewed as determined by an adversarial 
{\em scheduler}. 

In general, given a scheduler $S$  and a  set of robots 
$\R$, an {\em activation  sequence}  of $\R$ by (or, under) $\cal S$
  is  an infinite sequence  $E= \langle R_1, R_2, \ldots, R_i, \ldots \rangle$,
where   $R_i\subseteq \R$ denotes the non-empty set of robots
activated in round $i$,  satisfying   the {\em fairness constraint}:
$ (\forall r\in \R,\ \exists i : r\in R_i)\   \textbf{and}\  (\forall i \geq 1,  r\in R_i  \Rightarrow  \exists j>i : r\in R_j)$.  Let ${\cal E}({\cal S},\R)$ denote the set of all activation sequences  of $\R$ by $\cal S$; this set specifies the adversarial power of the scheduler.

The fairness constraint provides a temporal cost measure of robots' computations  in terms of (number of) successive sequences of activations of the robots, called \emph{epochs}: the first epoch starts with the first activation in the system;  each epoch ends as soon as all robots have been activated and they have completed their Look-Compute-Move cycle; the next epoch starts with the subsequent activation.

In the {\em synchronous} class of schedulers ({\sc Synch}),
time is logically divided into   \emph{rounds}.  
In each round, the scheduler activates a non-empty subset of the robots, and they perform each phase of their  \LCM\ cycle simultaneously, in perfect synchrony. 

We shall denote by \ssy\ the synchronous  scheduler, traditionally called {\em semi-synchronous},  without any other restriction (and, thus, with the strongest adversarial power in this class).
The special  {\em fully-synchronous} scheduler  \fsync,  which activates every robot  in every round (i.e., a round coincides with an epoch) corresponds to  further restricting   the activation sequences by:   
$\forall i\geq 1, R_i=\R$.

 The class of {\em sequential}  schedulers {\sc Sequential} $\subset$ {\sc Synch}  is defined  
 by   the restriction:  
 $\forall i\geq 1, |R_i|=1$,
that is, only one robot is activated in each round.
We shall denote by $\cal SEQ$  the sequential  scheduler  without any other restriction.

\subsection{Pattern Formation Problems}

A {\em configuration} $C(t)$ is the \emph{multiset} of the locations of the $n$ robots  at time $t$. Let $|Q(t)|$
be the corresponding set of unique points occupied by the robots at time $t$; obviously,
$|Q(t)| \leq |C(t)|$.

The \emph{Smallest Enclosing Circle}  of $C(t)$ is the  circle of smallest radius such that every point of $Q(t)$ lies on the circle or in its interior; it is denoted  by $SEC(Q(t))$. 
Let $\rho(t)$  and $c(Q(t))$ denote  the radius and the center of  $SEC(Q(t))$, respectively. 
In the following, should no ambiguity arise, the temporal indicator $t$ shall be omitted (ref. Fig.~\ref{fig:lambdastar}(a)). Furthermore, $c(Q)$ will be also denoted simply by $O$.

Since  robots 
 can only observe the positions of others and  move,
 in the ${\mathcal {OBLOT}}$ model,  a {\em problem} to be solved 
 is typically expressed in terms of  a   {\em temporal geometric predicate},
  which  the configuration   formed by the robot positions
must satisfy from some time on.

A {\em pattern}  $\hat{P}= \{\hat{p}_1, \hat{p}_2, \dots, \hat{p}_k\}\subset \mathbb R^2$  is a set of distinct points. 
A pattern $P'$ is said to be \emph{isomorphic} to a pattern $P''$ if $P''$ can be obtained by a combination
of translation, rotation, reflection and scaling of pattern $P'$.  We shall denote by ${\cal B}(\hat{P})$ the (infinite) set of patterns isomorphic to $P$.
We say the robots  have {\em formed} a  pattern $\hat{P}$ at time $t$, if the  configuration $C(t)$ is such that
$Q(t)\in {\cal B}(\hat{P})$.

The class of {\em Pattern Formation} problems is quite large, not only because of the different types of geometric patterns that can be formed, many types requiring special ad-hoc techniques for their formation (if at all possible),  but also for the scope of the requirements.

The most general problem in this class is   {\tt \Uni\  Pattern Formation}  ({\tt UPF})   that requires the robots, executing the same algorithm and  starting from any arbitrary initial configuration $C(0)$,
to form  any arbitrary pattern $\hat{P}$,   
given as their sole input and to no longer move.
It is defined  by the  
predicate:
\begin{align*}
{\tt UPF}& \equiv \forall   \hat{P}\subset \mathbb R^2, \forall n \geq |\hat{P}|,  \forall C\in (\mathbb R^2)^n,\ \\ & (C(0)=C) \Rightarrow \exists t\geq 0 :  (Q(t)\in {\cal B}(\hat{P}))\  \textbf{and}\ (\forall t'>t, C(t') = C(t)).
\end{align*}
\noindent where $(\mathbb R^2)^n$ denotes all possible set of $n$ points.

A particular instance is the  problem of forming  the pattern consisting of a single point, i.e., $|\hat{P}| =1$. %
This problem is special because it corresponds to the important coordination problem, called  {\tt Gathering} or {\tt Rendezvous}, of having all  robots meet at the same location, not fixed in advance (see \cite{Flocc19} for a recent survey);
it is defined by the
predicate: 
\begin{align*}
   {\tt Gather}&\equiv \forall n\in \mathbb Z^+ ,  \forall C\in (\mathbb R^2)^n,\  \\& (C(0)=C) \Rightarrow \exists t\geq 0 :  (|Q(t)|=1)\ \textbf{and}\   (\forall t'>t, Q(t') = Q(t)), 
\end{align*}
 \noindent 
 Due to its particular nature, the {\tt Gathering} problem has been treated in the literature differently from all other cases of pattern formation where $|\hat{P}|>1$. Also in this paper we shall make the same distinction, and refer to the case
 $|\hat{P}|>1$ of  {\tt UPF} as  ${\tt UPF^*}$.

\section{Impossibility of ${\tt UPF^*}$ and Preliminaries}
In this section, we provide some preliminary results, discussion and notation necessary for the understanding of the subsequent sections.

First,  
we show 
that ${\tt UPF^*}$ is
 unsolvable  under the fully-synchronous scheduler  {\fsync} and, thus, under \async\ and \ssy.

Second, we give an intuitive overview of our solution to solve {\tt UPF}$^*$ under \seq.

Finally, we provide the basic notation necessary for the formalization of our approach.

\subsection{Impossibility of {\tt UPF}$^*$ under \fsync}\label{sec:impossibility}

We start by showing the impossibility result.

\begin{theorem}\label{ImpossibilityFSYNC}
   ${\tt UPF^*}$ is unsolvable  under {\fsync} by a set  
 of $n\geq 3$ robots, 
  even if the robots have strong multiplicity detection, share the same coordinate system,  their movements are rigid, and one of them is (internally and externally) distinguishable from all the others.
\end{theorem}

\begin{proof}
    Consider a set $\R = \{r_1, r_2, r_3\}$ robots satisfying all the theorem's conditions and let us consider an initial configuration $C(0)$ such that exactly two robots form a multiplicity, whereas the third robot occupies a point alone. Moreover, let $S$ denote the set of configurations where all robots occupy distinct locations.

    Any strategy $\mathcal{A}$, operating under the {\fsync} scheduler, clearly cannot separate the $2$ robots forming the multiplicity, because of their symmetric views, activations, and deterministic behaviors.
    Therefore, starting from $C(0)$ with a multiplicity point, the robots cannot form any pattern in $S$. 
\end{proof}

On the other hand, it is known that the special case of {\tt Gathering} is solvable under \fsync\ without any additional robot capability \cite{CohenP05}.

\subsection{General strategy for solving  {\tt UPF}$^*$ under \seq}
\label{general}

%
Motivated by the impossibility results on the resolution of  {\tt UPF}$^*$ under \fsync, we investigate the problem under \seq.
At a very high level of description, our pattern formation  algorithm works as follows. 
The robots first move so to create a suitable  configuration 
where the number of points occupied by the robots is greater or equal than the number of pattern points, 
and there is a consensus among them  on a mapping between those two sets of points. 
The robots then start to  form the pattern by filling the empty pattern points
  according to an agreed upon priority order for the points occupation, defined in such a way that the agreement is preserved.

 Clearly, the actual execution of this simple high level strategy is complicated by a variety of technical issues and special cases, which require careful consideration.

As a first observation, the case of $1<|\hat{P}|< 5$ requires a specific approach as the general strategy cannot be applied due to the possible small number of robots involved. This case will be considered separately in Section~\ref{sec:P5}.
  
When $|\hat{P}|\geq 5$, instead, the strategy followed by the robots generally consists of four consecutive 
stages: {\em Initialization}, {\em Leader Configuration}, {\em Partial Pattern Formation}, and {\em Finalization}. Depending on current state of the configuration, some of these stages might be skipped.

The goal of the {\em Initialization} stage is to 
create a configuration $C$, if not already available, where the number $|Q|$ of distinct locations occupied by the robots is at least 
as large as the number $|\hat{P}|$ of points in the pattern, i.e., $|Q|\geq k = |\hat{P}|$.

Once $|Q| \geq k$, or if it was already as such from the initial configuration, the activated robot considers the pattern at a scale such 
that $SEC(\hat{P})$ overlaps $SEC(Q)$, i.e., their respective radii have the same length.

In the {\em Leader Configuration} stage, the goal is to create a specific map, called \emph{joint configuration}, 
that shows both the points occupied by the robots and the points of the pattern. 

Subsequently, in the {\em Partial Pattern Formation} stage, the pattern points are filled sequentially by the robots
through 
 movements 
along radial detours, that avoid  overpassing any robot. This stage lasts until only one robot position does not coincide with a pattern point.

When all but one pattern point are occupied, and all the robots not located on a pattern point are lying on a particular segment,  Procedure {\last} is invoked  for the {\em Finalization} stage. In this stage, all  the robots  reach their final destination and the pattern is formed.

Before describing the details of the algorithm, we need to introduce some important terminology and concepts which we will   employ  in the solution of  ${\tt UPF^*}$  under $\mathcal{SEQ}$
 without  additional robots' capabilities.

\subsection{Geometric Preliminaries}
\label{preliminaries}

\begin{table}[ht]
\centering
\renewcommand{\arraystretch}{1.2}
\begin{tabular}{ll}
\toprule
\textbf{Symbol} & \textbf{Description} \\
\midrule
$\mathcal{R}$ & Set of robots, $\mathcal{R}=\{r_1,\dots,r_n\}$ \\
$n$ & Number of robots, $n = |\mathcal{R}|$ \\
$\mathbb{R}^2$ & Euclidean plane \\
$C(t)$ & Configuration (multiset of robot positions) at time $t$ \\
$Q(t)$ & Set of distinct points occupied by robots at time $t$ \\
$|Q(t)|$ & Number of distinct occupied positions at time $t$ \\
$SEC(Q)$ & The smallest enclosing circle of $Q$ \\
$O$ & Center of $SEC(Q)$ \\
$\rho$ & Radius of $SEC(Q)$ \\
$\hat{P}$ & Target pattern, a finite set of points in $\mathbb{R}^2$ \\
$k$ & Number of points in the pattern, $k = |\hat{P}|$ \\
$\theta(a,b)$ & Angle at $O$ between points $a$ and $b$ \\
$\hat\mu$ & The smallest among the clockwise and counter-clockwise angular sequences\\
$\lambda(Q)$ & Leader angular sequence induced by $Q$ \\
${\cal B}(\hat{P})$ & Set of patterns isomorphic to $\hat{P}$ \\
$\delta$ & Minimum distance traveled by a robot if the destination is farther\\
\bottomrule
\end{tabular}
\caption{Summary of Notation}
\label{tab:notation}
\end{table}

Let $S\subset \mathbb R^2$  be a finite set of $l > 2$ points where no two points are \emph{co-radial},\footnote{Two points of a set $S$ are co-radial if they reside on the same radius of the $SEC(S)$.} and let  $c(S)$ be the center of $SEC(S)$.

The positions of the points of $S$, in the clockwise order, around  $c(S)$  induce a cyclic order $\psi $ among the points in $S$.  
Let $\psi(S) = (s_1, s_2, \dots, s_l)$ be the corresponding cyclic sequence of the points, 
with $s_1$ chosen arbitrarily, and let us  
denote by $\theta(s_i, s_{i+1})$ the angle, with respect to $c(S)$, between the two consecutive points $s_i$ and $s_{i+1}$ (ref. Fig.~\ref{fig:lambdastar}(b)).

Let $\theta(S) = \{\theta(s_i, s_{i+1}), \theta(s_{i+1}, s_{i+2}), \dots, \theta(s_{i-1}, s_{i})\}$ be an \emph{angular sequence} starting at $s_i$ induced on $S$.  
 
We now consider a set $S'$ containing either only robot positions $Q$ or both $Q$ and pattern points $\hat{P}$.
In the latter case,
we consider a suitable overlap (to be specified later) of $Q$ and $\hat{P}$, such that
$SEC(Q) = SEC(\hat{P})$.

If there are co-radial points (possibly robot positions or pattern points), they are all discarded from $S'$ but the most external one.
In this way, $S'$ does not contain co-radial points, $SEC(S') =SEC(Q) = SEC(\hat{P})$, and $|S'|=l'>2$.
An angular sequence $\theta(S') = (\theta_1, \dots,\theta_{l'})$ is a \emph{leader angular sequence}, denoted by $\lambda(Q)$ or $\lambda(Q\cup \hat{P})$, if \textit{(i)} $\theta_1 < \min\{\theta_2, \dots, \theta_{l'}\}$, and \textit{(ii)} the two adjacent points $s_i$ and $s_{i+1}$ defining $\theta_1$ belong to $Q$, with $s_i$ inside $SEC(Q)$ and $s_{i+1}$ on $SEC(Q)$. 

Without loss of generality, if $S'$ admits a leader angular sequence, then the rotational direction to overlap the radius containing $s_i$ with the radius containing $s_{i+1}$ spanning angle $\theta_1$, is considered to be clockwise. Therefore, 
a configuration $C(t)$ (possibly overlapped with a pattern $\hat{P}$), inducing the set $S'$ that admits a leader angular sequence, is asymmetric (apart for possible multiplicities) and all the robots can agree on the chirality of the system.

When $S'$ admits 
a leader angular sequence, we define the \emph{radiangular} distance between two points as follows.
Let $A$ and $B$ be two points that make the 
clockwise (according to the leader angular sequence)
angle $\theta = \angle AOB$ at $O$, the center of the $SEC$. 
The radiangular distance is defined by the tuple \textit{(i)} $(\theta, |\overline{OA}| + |\overline{OB}|)$, if $\theta > 0$, \textit{(ii)} $(0, |\overline{AB}|)$, otherwise. 
When $A$ and $B$ are not co-radial, the path of a robot from $A$ to $B$ consists of moving along $\overline{AO}$ to reach the center $O$, then along $\overline{OB}$ to reach $B$. 
When $A$ and $B$ are co-radial, the path is simply  
$\overline{AB}$. 
In both cases, the defined path is called a \emph{radial detour}.

Let us now consider the set of pattern points $\hat{P}$, with $|\hat{P}|\geq 5$. 
Two pattern points $p_i$ and 
$p_{i+1}$ are said to be \emph{adjacent} when:
\begin{itemize}
    \item they are co-radial and no other pattern point is in between them;
    \item they are not co-radial, there are no other pattern points on the segments $\overline{O\hat{p}_i}$ and $\overline{O\hat{p}_{i+1}}$, and there is no other pattern point inside the sector defined by angle $\theta(\hat{p}_{i}, \hat{p}_{i+1})$ calculated in the clockwise order.
\end{itemize}

Let $\alpha^{(i)}(\hat{P})=$ 
$\{(\theta(\hat{p}_{i}, \hat{p}_{i+1}), |\overline{O\hat{p}_i}|, |\overline{O\hat{p}_{i+1}}|)$, 
$(\theta(\hat{p}_{i+1}, \hat{p}_{i+2}),|\overline{O\hat{p}_{i+1}}|,|\overline{O\hat{p}_{i+2}}|)$,
$\dots$, 
$(\theta(\hat{p}_{i-1}, \hat{p}_{i}),
 |\overline{O\hat{p}_{i-1}}|, |\overline{O\hat{p}_{i}}|)\}$ 
be the set of triples determining what we call the \emph{clockwise pattern sequence} induced on $\hat{P}$ starting at an arbitrary $\hat{p}_i$. 
Similarly, let $\beta^{(i)}(\hat{P})=$
$\{(\theta(\hat{p}_i, \hat{p}_{i-1}), |\overline{O\hat{p}_i}|, |\overline{O\hat{p}_{i-1}}|)$,
$(\theta(\hat{p}_{i-1}, \hat{p}_{i-2}), |\overline{O\hat{p}_{i-1}}|, |\overline{O\hat{p}_{i-2}}|)$,
$\dots$, 
$(\theta(\hat{p}_{i+1}, \hat{p}_{i}), |\overline{O\hat{p}_{i+1}}|, |\overline{O\hat{p}_{i}}|)\}$ be the \emph{counter-clockwise pattern sequence}.
We shall denote by $\hat{\mu}$ the lexicographical smallest of all pattern sequences $\alpha^{(i)}(\hat{P})$ and  $\beta^{(i)}(\hat{P})$, $1\leq i\leq k$.
Note that, if $\hat{P}$ admits symmetries, then there is more than one index determining the same $\hat{\mu}$.

Table~\ref{tab:notation} summarizes the notations used throughout the paper.

\section{Solving {\tt UPF}$^*$ : $|\hat{P}|\geq 5$} 
\label{sec:P4}
%
In this section, we study the \texttt{\Uni\ Pattern} \texttt{Formation} problem under the $\mathcal{SEQ}$ scheduler in the case $|\hat{P}|\geq5$,
and prove that it is solvable even if the robots have no additional capabilities.
We do so by presenting a solution algorithm  under the $\mathcal{SEQ}$ scheduler when the robots have no multiplicity detection, no agreement on coordinate systems nor chirality, their movements are non-rigid,  $|{\cal R}|$ is unknown, and $|\hat{P}| \geq 5$. The case where $1< |\hat{P}| < 5$ will be addressed in the next section, since this case includes configurations with a very small number of robots, which makes our general strategy not applicable.

\subsection{Algorithm {\spf}}

Our solution to the Pattern Formation problem is provided by  Algorithm  {\spf}, which in turn exploits five procedures, determining the four 
stages described above: {\sepa} (in \emph{Initialization}), {\overl} and {\lead} (in \emph{Leader Configuration}), {\occ} (in \emph{Partial Pattern 
Formation}) and {\last} (in \emph{Finalization}); they are individually presented and described next, together with their pseudocodes. The correctness is presented in Section \ref{sec:complexSPF} and the flowchart representing the execution of the algorithm is presented in Fig. \ref{fig:flowchart}.

\usetikzlibrary{shapes.geometric, arrows, calc}
\tikzstyle{procedure} = [rectangle, rounded corners, minimum width=3cm, minimum height=1cm,text centered, draw=red]
\tikzset{ellipse node/.style={draw=gray, ellipse, minimum width=2cm, minimum height=1cm, thick}}
\tikzstyle{stage} = [rectangle, minimum width=3cm, minimum height=1cm, text centered]
\tikzstyle{decision} = [diamond, minimum width=3cm, minimum height=1cm, align=center, draw=blue]
\tikzstyle{arrow} = [thick,->,>=stealth]

\newcommand{\perparrow}[4]{
    \path ($(#1)!0.5!(#2)$) coordinate (mid);
    \path let \p1=(#1), \p2=(#2) in 
        coordinate (turn) at (\x2,\y1);
    \draw [arrow] (#1) -- node[#4] {#3} (turn) -- (#2);
}

\begin{figure}[t]
\centering
\resizebox{0.95\columnwidth}{!}{
\begin{tikzpicture}[node distance=2cm]
\def\h{3}
\def\l{9}



\node (start) [ellipse node] at (0.2*\l, 3.0*\h) {START};
\node (initcheck) [decision] at (0.2*\l, 2.2*\h) {$|Q| \stackrel{?}{<} |\hat{P}|$};
\node (sepa) [procedure] at (0.6*\l, 1.5*\h) {{\sepa}()};
\node (lastcheck) [decision] at (-0.5*\l, 1.4*\h) {Special case?};
\node (last) [procedure] at (-1*\l, 1*\h) {{\last}()};
\node (overl) [procedure] at (0.2*\l, 1*\h) {{\overl}()};
\node (hasleader) [decision] at (0.2*\l, 0.2*\h) {Has leader\\sequence?};
\node (lead) [procedure] at (-0.3*\l, 0.6*\h) {{\lead}()};
\node (occ) [procedure] at (0.6*\l, -0.4*\h) {{\occ}()};
\node (end) [ellipse node] at (-1.0*\l, -0.4*\h) {END};

\draw [arrow] (start) -- (initcheck);
\perparrow{initcheck}{sepa}{YES}{above}
\perparrow{initcheck}{lastcheck}{NO}{above}
\perparrow{lastcheck}{last}{YES}{above}
\perparrow{lastcheck}{overl}{NO}{above}
\draw [arrow] (last) -- (end);
\draw [arrow] (sepa) -| (initcheck);
\draw [arrow] (overl) -- (hasleader);
\perparrow{hasleader}{lead}{NO}{above}
\perparrow{hasleader}{occ}{YES}{above}
\draw [arrow] (lead) |- (overl);
\draw [arrow] (occ) -| (lastcheck);

\end{tikzpicture}}
\caption{Flowchart of Algorithm {\spf} showing the four stages: {\em Initialization}, {\em Leader Configuration}, {\em Partial Pattern Formation}, and {\em Finalization}. The flow shows how an activated robot processes the current configuration and makes movement decisions. { Note that, in the pseudo-code, the test on ``Special case?" is actually performed inside Procedure {\last}.}}
\label{fig:flowchart}
\end{figure}
\subsubsection{Overall Structure}

In this section, we describe  the overall structure of the resolution algorithm {\spf} from the point of view of the robot executing it.
An activated robot $r$, obtains a snapshot $Q$ in the {\em Look} phase, and then it executes Algorithm  {\spf} in its {\em Compute} phase. 
Let $q\in Q$ be the position of robot $r$.

The first operation the robot performs is to check if the number $|Q|$ of robots currently on distinct locations is sufficient to form the 
pattern. If $|Q| < k$, then robot $r$ executes Procedure {\sepa}, to determine an appropriate destination point, and terminates this execution. 
Otherwise, it proceeds to the next step to scale the pattern $\hat{P}$ in such a way that the radius 
of $SEC(\hat{P})$ has the same length $\rho$ of the radius of $SEC(Q)$ (ref. Fig.~\ref{fig:lambdastar}(d)).

Whenever $|Q| \geq k$, the activated robot checks if the current configuration $Q$ falls in one of the special cases, i.e., 
if it is already equivalent to the target pattern or the robot positions not coinciding with a specific 
pattern point are all located at $\overline{O\hat{p}_k}$.
In the former case, i.e., if the configuration already satisfies the pattern, no robot moves. In the latter, Procedure {\last} is invoked to finalize the formation of the pattern.

If the configuration does not correspond to   a special case,   Procedure {\last} returns $null$, and  the activated robot computes a ``joint configuration", denoted by $\map$,  to properly overlap the pattern on the current configuration (Procedure {\overl}).

If a joint configuration cannot be computed, i.e., $\map=null$, then there must be at least a robot that can 
move toward $O$ (provided it is not responsible for $SEC(Q)$ and it has a free radial path) by means of Procedure {\lead}, or there must be a robot 
located at $O$. In this latter case, again Procedure {\lead} is executed to create a leader configuration with a leader angular sequence.

Once $Q$ admits a leader angular sequence, the activated robot 
overlaps $Q$ and $\hat{P}$ in a specific way creating the joint configuration $\map$, and then it checks if $\map$ admits a leader angular sequence. If not, Procedure {\lead} is executed to create 
a leader angular sequence in $\map$. Otherwise, 
Procedure {\occ} is executed. The goal of this 
Procedure is to move a specific robot 
to the highest priority empty pattern point $p_l$. 

Once all robot positions but the lowest priority one coincide with the highest priority pattern points, the robots are in one of the special cases described above. 

\begin{algorithm}[ht]
\KwIn{$\hat{P}$, $Q$} 
\KwOut{Destination of the robot}
$q \gets$ activated robot position\;
\eIf{$|Q| < |\hat{P}| = k$}{
    destination $\gets$ {\sepa}()\;\label{UPFseparate}
}{  Scale $\hat{P}$ such that radius of $SEC(\hat{P})$ is $\rho$\;
    
    $res \gets$ {\last}()\;\label{UPFlast}
    \eIf{$res \neq null$}{destination $\gets res$\;}{\label{spf:else}
    $\map \gets $ {\overl()}\;\label{UPFoverlap}
    \eIf{$\map = null$}{
            destination $\gets$ {\lead}()\;\label{UPFleader1}
    }
    {
        \eIf{$\map$ has a leader angular sequence}{
            destination $\gets$ {\occ}()\;\label{UPFoccupy}
        }
        {
          destination $\gets$ {\lead}()\;\label{UPFleader2}  
        }
    } 
}
}

\caption {\spf}\label{algo:upf}
\end{algorithm}


\subsubsection{Procedure {\sepa}}
This procedure consists of a very simple operation.
The activated robot $r$ computes a point $q'$ along its own $x$-axis at one unit distance. If $\overline{qq'}$ 
does not contain other robots, $r$ selects as its destination $q'$. Otherwise, $r$ computes the position 
of its closest robot $q''$ along its own $x$-axis, and selects as its destination the midpoint of $\overline{qq''}$.

\begin{algorithm}[ht]
\SetAlgorithmName{Procedure}{List of Procedures}{List of Procedures}
\caption{{\sepa}()}\label{algo:sepa}
$q' \gets$ a point at one unit distance from $q$ along the $x$-axis\;
\eIf{$\overline{qq'}$ does not contain other robots}{
    $z \gets q'$\; }
    {
        $q'' \gets$ robot position closer to $q$ along the $x$-axis\;
        $z \gets$ midpoint of $\overline{qq''}$\;
    }

    \Return $z$\;
\end{algorithm}


\subsubsection{Procedure {\last}}\label{sec:last}
As previously described, Procedure {\last} is executed to manage some special cases.
Once the activated robot $r$ has verified that $|Q| \geq k$, it starts by checking if the current configuration  is already the target pattern, i.e.,  $Q \in {\cal B}(\hat{P})$.
Robot $r$ checks this by rotating and reflecting the pattern as follows. 
Let $\hat{p}_1$ be a pattern point on $SEC(\hat{P})$.
Consider all the robot positions $(q_1, \dots, q_s) = SEC(Q)\cap Q$ in the cyclic order of some $\psi(S')$. Let $q_i$ be a robot position in $(q_1, \dots, q_s)$.
The robot $r$ overlaps the pattern point $\hat{p}_1$ with each of the robot positions $q_i$ ``one by one'' by 
reflecting the
pattern (across the radius $\overline{O\hat{p}_1}$). If one of these transformations
exactly overlaps all the points of $Q$ with all the points of $\hat{P}$, it follows that $Q \in {\cal B}(\hat{P})$, and $r$ does not move. 
In the worst case, if all the robot positions are on the $SEC(Q)$, then $r$ has to check at most $2k$ transformations ($k$ rotations and $k$ reflections). 
Note that  the case when the $SEC(Q)$ contains more than $k$ robot positions  is not handled by this procedure.

If $|Q| \geq k$ and $Q \notin {\cal B}(\hat{P})$, 
the activated robot checks if the configuration falls into the special cases ``last one''. 
These cases occur when pattern points $\hat{p_i}$, $1\leq i< k$, 
are occupied and the only robot positions not on any pattern point are located at $\overline{O\hat{p}_k}$, with 
$\hat{p}_k$ being the lowest priority pattern point.  

The lowest priority pattern point $\hat{p}_k$ is defined by any lexicographical smallest pattern sequence $\hat{\mu}$.
Once a robot located at $\overline{O\hat{p}_k}$ is activated, it moves toward $\hat{p}_k$, 
if not yet there. This case handles the cases  
where all the pattern points are occupied, except for the last one 
$\hat{p}_k$, and the cases where all the pattern points are occupied and there are still robots moving toward $\hat{p}_k$. In both   scenarios, the only robot positions not on pattern points, are  on the segment $\overline{O\hat{p_k}}$.

\begin{algorithm}[ht]
\SetAlgorithmName{Procedure}{List of Procedures}{List of Procedures}
\caption{{\last}()}\label{algo:last}
\If{$|Q| \geq k$}{
Compute at most $2k$ transformations $\mathcal{T}$ of the pattern $\hat{P}$ such that pattern point $\hat{p}_1$ overlaps 
with any $q_i \in SEC(Q)$\;\label{last:T}
    For each $\tau \in \mathcal{T}$, let $\hat{p}_k \in \hat{P}$ be the lowest priority pattern point defined by any lexicographical smallest pattern sequence $\hat{\mu}$\;
\eIf{$\exists$ $\tau \in \mathcal{T}$ such that $Q \in {\cal B}(\hat{P})$}{
        \Return $q$\;\label{last:F}
    }
    {
    \If{$\exists$ $\tau \in \mathcal{T}$ such that $q \in \overline{O\hat{p}_k}$ and
     $(\hat{P}\setminus Q = \hat{p}_k$ $\vee$ $\hat{P}\setminus Q = \emptyset)$ $\wedge$
     $Q \setminus \hat{P} = (Q \cap \overline{O\hat{p_k}})$  
     \label{last:cases}} 
    {
        \Return $\hat{p}_k$\label{last:move}}
    }}
\Return $null$\;
\end{algorithm}


\subsubsection{Procedure {\overl}}\label{sec:overlap}

When the activated robot $r$ determines that $|Q| \geq k$ and that the configuration does not concern the special cases managed by Procedure 
{\last} (i.e., call {\last()} returns $null$), it checks if $Q$ has a leader angular sequence $\lambda(Q)$. 
If this is the case,
$r$ proceeds to overlap the pattern points on the current configuration and subsequently achieve a joint configuration $\map$. 
\begin{figure}[]
    \centering
    \subfloat[A configuration $C(t)$]{
        \resizebox{0.33\textwidth}{!}{
            \begin{tikzpicture}
                \def\innerradius{2} 
                \def\outerradius{3} 

                \draw[dashed] (0,0) circle (\outerradius);
                \draw[draw = none] (0,0) circle (\outerradius+0.5);
                
                \foreach \r/\a [count = \i] in {\outerradius/10, \outerradius/45, 1.4/55, \outerradius/100, 0.8/100, 0.8/140, \outerradius/160, 0.5/200, 2.2/240, 1.8/290, \outerradius/310}{
                    \fill (\a:\r) circle (2pt);
                    \node[fill=white, circle, inner sep=0pt, auto] at (\a:\r+0.3) {$q_{\i}$};
                }
            \end{tikzpicture}
        }
    }
    \subfloat[{Angular sequence\\ of $S'$}]{
        \resizebox{0.33\textwidth}{!}{
            \begin{tikzpicture}
                \def\innerradius{2} 
                \def\outerradius{3} 

                \draw[dashed] (0,0) circle (\outerradius);
                \draw[draw = none] (0,0) circle (\outerradius+0.5);
                
                \foreach \r/\a [count = \i] in {\outerradius/10, \outerradius/45, 1.4/55, \outerradius/100, 0.8/140, \outerradius/160, 0.5/200, 2.2/240, 1.8/290, \outerradius/310}{
                    \fill (\a:\r) circle (2pt);
                    \draw (0,0) -- (\a:\r);
                    \draw[dashed] (\a:\r) -- (\a:\outerradius);
                    \node[fill=white, circle, inner sep=0pt, auto] at (\a:\r+0.3) {$s_{\i}$};
                }
                \foreach \b/\i [count=\j from 2] in {25/1, 50/2, 80.0/3, 120/4, 150/5, 180/6, 220/7, 265/8, 300/9}{
                    \node[rotate = \b] at (\b:2) {$\theta(s_{\i}, s_{\j})$};
                }
                \node[rotate = -20] at (-20:2) {$\theta(s_{10}, s_{1})$};
            \end{tikzpicture}
        }
    }
    \subfloat[Smallest unique angle in $C(t)$]{
        \resizebox{0.33\textwidth}{!}{
            \begin{tikzpicture}
                \def\innerradius{2} 
                \def\outerradius{3} 

                \draw[dashed] (0,0) circle (\outerradius);
                \draw[draw = none] (0,0) circle (\outerradius+0.5);
                
                \foreach \r/\a [count = \i] in {\outerradius/10, 3/45, 1.4/55, 0.8/100, \outerradius/100, 0.8/140, 0.5/200, \outerradius/160, 2.2/240, 1.8/290, \outerradius/310}{
                    \fill (\a:\r) circle (2pt);
                    \draw (0,0) -- (\a:\r);
                    \draw[dashed] (\a:\r) -- (\a:\outerradius);
                }
                \foreach \b/\i [count=\j] in { 50/2, 25/1, -20/9, 300/8, 265/7, 200/6, 150/5, 120/4, 80.0/3}{
                }
                \node[rotate=-40] at (50:\outerradius+0.3) {$\xi$};

            \end{tikzpicture}
        }
    }\\
    \subfloat[$\hat{P}$]{
        \resizebox{0.33\textwidth}{!}{
            \begin{tikzpicture}
                \def\innerradius{2}
                \def\outerradius{3}

                \draw[dashed] (0,0) circle (\outerradius);
                \draw[draw = none] (0,0) circle (\outerradius+0.5);
                \foreach \r/\a [count  = \i] in {1/45, \outerradius/45, 0.5/90, \outerradius/135, 1/180, \outerradius/225, 0.5/270, 1/300, \outerradius/315}{
                    \draw[blue, thick] (450-\a:\r) circle (2pt);
                    \node at (450-\a:\r+0.3) {$\hat{p}_\i$};
                }
                    
            \end{tikzpicture}
        }
    }
    \subfloat[Circular decomposition of $\hat{P}$]{
        \resizebox{0.33\textwidth}{!}{
            \begin{tikzpicture}
                \def\innerradius{2}
                \def\outerradius{3}

                \draw[dashed] (0,0) circle (\outerradius);
                \draw[dashed] (0,0) circle (1);
                \draw[dashed] (0,0) circle (0.5);
                \draw[draw = none] (0,0) circle (\outerradius+0.5);
                \foreach \r/\a [count  = \i] in {1/45,\outerradius/45,  0.5/90, \outerradius/135, 1/180, \outerradius/225, 0.5/270, 1/300, \outerradius/315}{
                    \draw[blue, thick] (450-\a:\r) circle (2pt);
                    \node at (450-\a:\r+0.3) {$\hat{p}_\i$};
                }
                \draw[-{Stealth[red]}, rotate = 90] (0,0) -- node[midway, left] {$\rho_1$} (\outerradius,0);
                \draw[-{Stealth[red]}, rotate=-50] (0,0) -- node[midway, below] {$\rho_2$} (1,0);
                \draw[-{Stealth[red]}, rotate = 20] (0,0) -- node[midway, above] {$\rho_3$} (0.5,0);
                    
            \end{tikzpicture}
        }
    }\subfloat[$\hat{\mu}$ of $\hat{P}$]{
        \resizebox{0.33\textwidth}{!}{
            \begin{tikzpicture}
                \def\innerradius{2}
                \def\outerradius{3}

                \draw[dashed] (0,0) circle (\outerradius);
                \draw[draw = none] (0,0) circle (\outerradius+0.5);
                \foreach \r/\a [count  = \i] in {1/45, \outerradius/45, 0.5/90, \outerradius/135, 1/180, \outerradius/225, 0.5/270, 1/300, \outerradius/315}{
                    \draw[blue, thick] (450-\a:\r) circle (2pt);
                    \draw (0,0) -- (450-\a:\r);
                    \draw[dashed] (450-\a:\r) -- (450-\a:\outerradius);
                    \node at (450-\a:\r + 0.2) [shift={(450-\a+90:0.2)}]{$\hat{p}_\i$};
                }
                \foreach \b/\i/\theta/\op/\opNext in {
    40/1/0/0.33/1,      
    67.5/2/45/1/0.17,  
    112.5/3/45/0.17/1,    
    157.5/4/45/1/0.33,    
    202.5/5/45/0.33/1,    
    247.5/6/45/1/0.17,    
    285/7/30/0.17/0.33,   
    307.5/8/15/0.33/1,    
    0/9/90/1/0.33           
}{
    \node[rotate={(450-\b)}, font=\tiny, scale=0.8] at (450-\b:2.2) {$(\theta^\circ, \op,\opNext)$};
}

                 \foreach \r/\a [count = \i] in {\outerradius/45, \outerradius/135, \outerradius/225, \outerradius/315}{
                    
                }
            \end{tikzpicture}
        }
    }\\
    \subfloat[Joint configuration $\map(t)$]{
        \resizebox{0.33\textwidth}{!}{
            \begin{tikzpicture}
                \def\innerradius{2}
                \def\outerradius{3}
                \def\offset{5}
                \def\rotation{455}
                \draw[dashed] (0,0) circle (\outerradius);
                \draw[draw = none] (0,0) circle (\outerradius+0.5);
                \foreach \r/\a [count  = \i] in {1/45, \outerradius/45,  0.5/90, \outerradius/135, 1/180, \outerradius/225, 0.5/270, 1/300, \outerradius/315}{
                    \draw[blue, thick] (\rotation-\offset-\a:\r) circle (2pt);
                    \draw (0,0) -- (\rotation-\offset-\a:\r);
                    \draw[dashed] (\rotation-\offset-\a:\r) -- (\rotation-\offset-\a:\outerradius);
                }
                \node[fill=white, circle, inner sep=0pt, left] at (55:1.5 + 0.3) {$u'$};
                
                \foreach \r/\a [count = \i] in {\outerradius/10, 3/45, 1.4/55, 0.8/100, \outerradius/100, 0.8/140, 0.5/200, \outerradius/160, 2.2/240, 1.8/290, \outerradius/310}{
                    \fill (\a:\r) circle (2pt);
                    \draw (0,0) -- (\a:\r);
                    \draw[dashed] (\a:\r) -- (\a:\outerradius);
                }
                \draw[red, thick] (55:1.4) circle (2pt);
                \draw[dash dot] (0,0) -- (55:1.4);
            \end{tikzpicture}
        }
    }
    \subfloat[Leader configuration]{
        \resizebox{0.33\textwidth}{!}{
            \begin{tikzpicture}
                \def\innerradius{2}
                \def\outerradius{3}
                \def\offset{5}
                \def\rotation{455}
                \draw[dashed] (0,0) circle (\outerradius);
                \draw[draw = none] (0,0) circle (\outerradius+0.5);
                \foreach \r/\a [count  = \i] in {1/45, \outerradius/45, 0.5/90, \outerradius/135, 1/180, \outerradius/225, 0.5/270, 1/300, \outerradius/315}{
                    \draw[blue, thick] (\rotation-\offset-\a:\r) circle (2pt);
                }
                \foreach \r/\a [count = \i] in {\outerradius/10, 3/45, 1.4/55, 0.8/100, \outerradius/100, 0.8/140, 0.5/200, \outerradius/160, 2.2/240, 1.8/290, \outerradius/310}{
                    \fill (\a:\r) circle (2pt);
                }
                \draw (0,0) -- (55:1.4);
                \draw[dashed] (0,0) -- (55:3);
                \draw (0,0) -- (45:3);
                \draw[-{Stealth[red]}] (55:1.4) arc (55:-2:2);
                \fill (55:1.4) circle (2pt);
                \node[fill=white, circle, inner sep=0pt, left] at (55:1.5+0.3) {$u'$};
                \node[fill=white, circle, inner sep=0pt, auto] at (40:3+0.3) {$q_j$};
            \end{tikzpicture}
        }
    }
    \subfloat[Ranking of pattern points]{
        \resizebox{0.33\textwidth}{!}{
            \begin{tikzpicture}
                \def\innerradius{2}
                \def\outerradius{3}
                \def\offset{5}
                \def\rotation{450}
                \draw[dashed] (0,0) circle (\outerradius);
                \draw[dashed] (0,0) circle (1);
                \draw[dashed] (0,0) circle (0.5);
                \draw[draw = none] (0,0) circle (\outerradius+0.5);
                \foreach \r/\a [count  = \i] in {\outerradius/45, \outerradius/225, \outerradius/135,\outerradius/315, 1/45,  1/180, 1/300, 0.5/90, 0.5/270}{
                    \draw[blue, thick] (\rotation-\offset-\a:\r) circle (2pt);
                    \node at (\rotation-\offset-\a:\r + 0.3) [shift={(450-\a+90:0.2)}] {$p_\i$};
                }
                \foreach \r/\a [count = \i] in {\outerradius/10, 3/40, 1.5/42.5, 1.4/60, 0.8/100, \outerradius/100, 0.8/140, \outerradius/160, 2.2/240, 1.8/290, \outerradius/310}{
                }
                \draw (0,0) -- (40:3);
                \draw[-{Stealth[red]}] (40:2) arc (42.5:-10:2);
            \end{tikzpicture}
        }
    }
    \caption{(a) A configuration $C(t)$ and $SEC(Q(t))$; (b) The subset of non co-radial points; (c) Unique smallest angle $\xi$ of $Q(t)$; (d) The pattern $\hat{P}$; (e) Circular decomposition $\Cir$ of the pattern $\hat{P}$; (f) The smallest angular sequence $\hat{\mu}$ of $\hat{P}$; (g) The joint configuration $\map(t)$; (h) The leader configuration and the induced clockwise direction starting from $u'$, the robot position inside the $SEC$ defining the unique smallest angle; (i) The ranking of the pattern points with $q_j = p_1$.}
    \label{fig:lambdastar}
\end{figure}

Let $\Cir= \{\cir_1, \cir_2, \dots, \cir_\sigma\}$ be the set of all concentric circles, in the decreasing order of their radii, containing points of $\hat{P}$ (ref. Fig.~\ref{fig:lambdastar}(e)), and let $\rho_j$ be the radius of $\cir_j$, $1\leq j \leq \sigma$;  thus $\cir_1 = SEC(\hat{P})$ and $\rho_1=\rho$ (ref. Fig.~\ref{fig:lambdastar}(f)). 
The objective of the robot is to 
rotate and/or reflect $\hat{P}$ until some points in the $SEC(Q)$ and $SEC(\hat{P})$, with specific properties, are co-located, i.e., they precisely overlap. This is done as follows. 
 
Let $\hat{\mu}$ be the lexicographical smallest pattern sequence of $\hat{P}$ and $\psi(\hat{P})$ be the corresponding cyclic 
order of points. 
Let $\hat{p}_i$ be the first pattern point on the $SEC(\hat{P})$ in the cyclic order $\psi(\hat{P})$. 
Let $q_j$ be the second robot position determining $\theta_1$ in $\lambda(Q)$.
The activated robot proceeds to overlap $Q$ and $\hat{P}$ so that $q_j$ overlaps with $\hat{p}_i$ and the order induced by $\hat{\mu}$ follows the clockwise direction induced by $\lambda(Q)$.

\begin{algorithm}[ht]
Let $\hat{p}_i$ be the first pattern point on the $SEC(\hat{P})$ in the cyclic order $\psi(\hat{P})$ determined by $\hat{\mu}$\;

\eIf{$Q$ admits a leader angular sequence}{
        Let $q_j$ be the
        second robot position determining $\theta_1$ in $\lambda(Q)$\;
        Overlap $Q$ and $\hat{P}$ in such a way that $q_j$ overlaps with $\hat{p}_i$, $Q$ is considered in the clockwise order induced by $\lambda(Q)$ and the order induced by $\hat{\mu}$ follows such clockwise direction\;
        \Return $\map$\;
        
}{
        \Return $null$\;
}
\SetAlgorithmName{Procedure}{List of Procedures}{List of Procedures}
\caption{{\overl()}}\label{algo:overl}
\end{algorithm}


\subsubsection{Procedure {\lead}}\label{sec:leader}
As input, Procedure {\lead} can get: a configuration $Q$ with no leader angular sequence if $\map = null$ or a joint 
configuration $\map$ (ref. Fig.~\ref{fig:lambdastar}(g)) if Procedure {\overl} has been executed from a configuration $Q$ admitting a leader angular sequence.

The objective of {\lead} is to first send a robot to $O$ if $O$ is empty, and then to create a leader configuration (ref. Fig.~\ref{fig:lambdastar}(h)). 
If $O$ is empty, then the robots that can move are the ones not responsible for the $SEC(Q)$ and 
that have a path with no other robot positions toward $O$, or if there is a joint configuration 
$\map$, the robot that also cannot move is the one responsible for the smallest angle $\theta_1$ if it has 
no other co-radial robots.

If, instead, $O$ is occupied by a robot, once such robot is activated,
it computes $\xi$, the smallest angle in the 
input configuration, and moves to $u'$ such that $|\overline{u'O}| = \rho/2$ and for a given $q_j \in SEC(Q)$, it forms $\angle u'Oq_j = \xi/3$.

\begin{algorithm}[ht]
\SetAlgorithmName{Procedure}{List of Procedures}{List of Procedures}
\eIf{there is no robot at $O$}{
    \If{$q$ is responsible for the $SEC(Q)$ or $\overline{Oq}$ contains other robot positions}{
        \Return $q$\;
    }
    \eIf{$\map \neq null$ and $q$ is responsible for $\theta_1$ and $q$ is not co-radial to another robot position}{
        \Return $q$\;
    } {
            \Return $O$\;            
        }
    }{
    \eIf{$q = O$}{
        \eIf{$\map=null$}{
            Let $q_j$ any robot on $SEC(Q)$\;
            $\xi \gets$ smallest angle in $\theta(S')$, with $S'$ induced by $Q$\;
        }{
            Let $q_j$ be the second robot determining $\theta_1$ in $\theta(S')$, with $S'$ induced by $\map$\;
            $\xi \gets$ smallest angle in $\theta(S')$, with $S'$ induced by $\map$\;
        }
        Let $u'$ such that $|\overline{Ou'}| = \rho/2$ and $\angle u'Oq_j = \xi/3$\;
        \Return $u'$\;
    }{
        \Return $q$\;
    }
}
\caption{{\lead()}}\label{algo:lead}
\end{algorithm}


\subsubsection{Procedure {\occ}}

Procedure {\occ} deals with the movement of robots to reach pattern points. 
Given a joint configuration $\map$ with leader angular sequence, the ranking of the pattern points can be established.
A robot is then chosen as the one to move, i.e., the \emph{walker}, provided it 
satisfies some conditions. The walker robot moves to the target pattern point via a 
radial detour.
Recall that $u'$ and $q_j$ are the robot positions defining 
the smallest angle $\theta_1$ with $q_j$ on the $SEC(\map)$. 
Let $q_j$ be located at $A$.
Now $\overline{OA}$ is the radius. 
Consider points $A_2, \ldots, A_\sigma$ be the projections of $A$ on the circles $\cir_2, \cir_3, \dots, \cir_\sigma$, respectively.
For each $\cir_i$, the pattern points are ranked in the increasing order starting from $A_i$ and moving following the clockwise direction induced by $\lambda(\map)$
until we reach $A_i$ again.
The rank of pattern points in $\cir_i$ is smaller than $\cir_{i+1}$ for any $i = 1, \dots, \theta - 1$. 
For $\cir_1$, the rank follows a different strategy. This is required in order to maintain the original 
$SEC$ during the execution of the algorithm. By Procedure {\overl}, we 
already have a robot at $q_j$, which is also the highest priority pattern point. We 
start at $q_j$ as $p_1$.
Consider $p_1'$ as the antipodal point of $p_1$, positioned on the $SEC$. If there is a pattern point at $p_1'$, then that becomes $p_2$. Otherwise, consider the two closest pattern points to $p_1'$ positioned on the $SEC$, going from $p_1'$ to $p_1$, one 
in the counterclockwise direction and the other one in the clockwise direction.
They respectively become $p_2$ and $p_3$.
The remaining pattern points on $\cir_1$ are assigned in the increasing order as usual.
Notice that, $p_1$ is the smallest rank pattern point and $p_k$ is the highest rank pattern point. Point $p_k$ is located on $\cir_\sigma$ (ref. Fig.~\ref{fig:lambdastar}(i)). 

\begin{algorithm}[ht]
$p_k\gets$ the lowest priority pattern point\;
\eIf{all pattern points are occupied}{$p_l \gets p_k$\;}{
$p_l \gets$ the highest priority empty pattern point\;}
Determine the highest priority free robot position $u$ not defining the $SEC(Q(t))$\;
\If{the radial detour from $u$ to $p_l$, and $\rho_l$, do not contain other robots}{
    $w \gets u$\;
}
\ElseIf{the radial detour from $u$ to $p_l$ contains another robot and $\rho_l$ does not}   
 {along the radial detour from $u$ to $p_l$, determine the robot position $v$ 
 with smallest radiangular distance from $p_l$\;
    $w \gets v$\;}
 
\Else 
 { 
    consider $\rho_l$ from the $SEC$ to $O$ and 
    compute the path from the first free robot position $v$ to $p_l$\;
    \eIf{$\overline{vp_l}$ does not contain other robot positions}{
        $w \gets v$\;    
    } {in $\overline{vp_l}$
    determine the closest robot position $v'$ to $p_l$ with free path\;
    $w \gets v'$\;}}

\eIf{$q = w$}{
    \eIf{$q$ and $p_l$ are co-radial}{
    \Return $p_l$\;
    }{
    \Return $O$\;
    }
}{
    \Return $q$\;
}
\SetAlgorithmName{Procedure}{List of Procedures}{List of Procedures}
\caption{{\occ}()}\label{algo:occ}
\end{algorithm}

\subsubsection{Defining the walker}\label{sec:walker}

The walker is a robot in the current configuration that will move to occupy an unoccupied pattern point. Before we define the walker,
we define an order among the robots. 
This order 
uses the radiangular distance 
from $p_1$ and we say that a robot is \emph{free} if it is not located at a pattern point.  
Let $r'$ and $r''$ be two robots that have radiangular distance from $p_1$ equal to $(\theta', x')$ 
and $(\theta'',x'')$, respectively. 
We say that $r'$ has higher priority than $r''$, if either $(i)$ $\theta' < \theta''$,  
$(ii)$ $\theta' = \theta'' \neq 0$ and $x' > x''$, 
or $(iii)$ $\theta' = \theta'' = 0$ and $x' < x''$.
In other words, a robot that is closer in angle to 
the radius $\overline{Op_1}$ has higher priority, and when two (or more) robots are on the same radius, the robot that is farther from the 
center has higher priority.

Now, consider the radial detour from the highest priority free robot to the highest priority empty  pattern point, and the radius $\rho_l$ where such a point lies. The following could happen:
\begin{itemize}
    \item If the radial detour and $\rho_l$ do not contain other robots, then the highest priority free robot becomes the walker;
    
    \item If the radial detour contains another robot and $\rho_l$ does not contain other robots, the robot with smallest radiangular distance from the target pattern point, along the computed radial detour, becomes the walker;

    \item Otherwise, if $\rho_l$ contains other robots, 
    compute the path from the highest priority free robot in $\rho_l$ to $p_l$:
     if this path is empty, then the free robot 
    becomes the walker, otherwise, in the computed path, the robot closest to $p_l$ becomes the walker.
\end{itemize}

Notice that the walker has always a free path to the target pattern point and that under these restrictions, the internal robot defining $\theta_1$ becomes the last walker since it has the largest radiangular distance from $p_1$.

Also, notice that once the pattern points fixing the $SEC(\hat{P})$ are occupied by robots, the 
remaining robots on the $SEC(Q)$ are free to move, since they will not be 
necessary to maintain the $SEC$ anymore. 

\subsubsection{Movement of a robot}
The movement of a robot always follows a radial detour. 
If a robot $r$ is determined as the walker and its target pattern point 
is located on a different radius from the one where it is currently located, $r$ first moves to the center, and then moves toward its target pattern point from the center. 
When $r$ is co-radial with its target pattern point, it moves directly to that point. By definition, the walker does not have any other robot on the radial detour, and hence it can move to its target pattern point unobstructed. In Fig.~\ref{fig:movement}, we show the movement of the walker to its corresponding destination.

\usetikzlibrary{decorations.markings}
\begin{figure}[H]
    \centering
    \subfloat[$q_1$ moves towards $p_2$]{
        \resizebox{0.33\textwidth}{!}{
            \begin{tikzpicture}[decoration={markings,
  mark=between positions 0.25 and 0.9 step 0.25 with {\arrow{stealth}}}]
                \def\innerradius{2}
                \def\outerradius{3}
                \def\offset{5}
                \def\rotation{455}
                \draw[dashed] (0,0) circle (\outerradius);
                \draw[draw = none] (0,0) circle (\outerradius+0.5);
                \foreach \r/\a [count  = \i] in {\outerradius/45,   \outerradius/225,\outerradius/135,\outerradius/315, 1/45,  1/180, 1/300, 0.5/90, 0.5/270}{
                    \draw[blue, thick] (\rotation-\offset-\a:\r) circle (2pt);
                    \node at (\rotation-\offset-\a:\r + 0.3) {$p_\i$};
                }
                \foreach \r/\a [count = \i] in {\outerradius/10, 3/45, 1.4/55, 0.8/100, \outerradius/100, 0.8/140, 0.5/200, \outerradius/160, 2.2/240, 1.8/290, \outerradius/310}{
                    \fill (\a:\r) circle (2pt);
                }
                \draw[dashed] (0,0) -- (45:3);
                \draw[-{Stealth[red]}] (45:2) arc (45:-10:2);
                \draw[postaction={decorate}] (10:3) -- (0,0);
                \draw[postaction={decorate}] (0,0) -- (225:3);
                \node at (10:3+0.3) {$q_1$};
            \end{tikzpicture}
        }
    } 
    \subfloat[$q_{10}$ moves towards $p_4$]{
        \resizebox{0.33\textwidth}{!}{
            \begin{tikzpicture}[decoration={markings,
  mark=between positions 0.25 and 0.9 step 0.25 with {\arrow{stealth}}}]
                \def\innerradius{2}
                \def\outerradius{3}
                \def\offset{5}
                \def\rotation{455}
                \draw[dashed] (0,0) circle (\outerradius);
                \draw[draw = none] (0,0) circle (\outerradius+0.5);
                \foreach \r/\a [count  = \i] in {\outerradius/45,  \outerradius/225,\outerradius/135,\outerradius/315, 1/45,  1/180, 1/300, 0.5/90, 0.5/270}{
                    \draw[blue, thick] (\rotation-\offset-\a:\r) circle (2pt);
                    \node at (\rotation-\offset-\a:\r + 0.3) {$p_\i$};
                }
                \foreach \r/\a [count = \i] in {\outerradius/220, 3/45, 1.4/55, 0.8/100, \outerradius/100, 0.8/140, 0.5/200, \outerradius/160, 2.2/240, 1.8/290, \outerradius/310}{
                    \fill (\a:\r) circle (2pt);
                }
                \draw[dashed] (0,0) -- (40:3);
                \draw[-{Stealth[red]}] (40:2) arc (42.5:-10:2);
                \draw[postaction={decorate}] (290:1.8) -- (0,0);
                \draw[postaction={decorate}] (0,0) -- (135:3);
                \node at (290:1.8+0.3) {$q_{10}$};
            \end{tikzpicture}
        }
    } 
    \subfloat[Robot at $u'$ moves towards $p_9$]{
        \resizebox{0.33\textwidth}{!}{
            \begin{tikzpicture}[decoration={markings,
  mark=between positions 0.25 and 0.9 step 0.4 with {\arrow{stealth}}}]
                \def\innerradius{2}
                \def\outerradius{3}
                \def\offset{5}
                \def\rotation{455}
                \draw[dashed] (0,0) circle (\outerradius);
                \draw[draw = none] (0,0) circle (\outerradius+0.5);
                \foreach \r/\a [count  = \i] in {\outerradius/45,  \outerradius/225,\outerradius/135,\outerradius/315, 1/45,  1/180, 1/300, 0.5/90, 0.5/270}{
                    \draw[blue, thick] (\rotation-\offset-\a:\r) circle (2pt);
                    \node at (\rotation-\offset-\a:\r + 0.3) {$p_\i$};
                    \fill (\rotation-\offset-\a:\r) circle (2pt);
                }
                \foreach \r/\a [count = \i] in {\outerradius/220, 3/45, 1.4/55, 0.8/100, \outerradius/100, 0.8/140, \outerradius/160, 2.2/240, 1.8/290, \outerradius/310}{
                }
                \fill (55:1.5) circle (2pt);
                \draw[dashed] (0,0) -- (45:3);
                \draw[-{Stealth[red]}] (45:2) arc (45:-10:2);
                \draw[postaction={decorate}] (55:1.5) -- (0,0);
                \draw[postaction={decorate}] (0,0) -- (180:0.5);
                \node[fill=white, circle, inner sep=0pt, left] at (55:1.5+0.4) {$u'$};
            \end{tikzpicture}
        }
    } 
    \caption{Movement of robots to occupy the pattern points in the order of priority.}
    \label{fig:movement}
\end{figure}


\subsection{Correctness and Complexity of {\spf}} \label{sec:complexSPF}


In this section, we show the correctness of the proposed Algorithm  {\spf} and discuss 
its complexity.
\ifthenelse{\boolean{useInline}}{}{
    The proofs are in Appendix~\ref{app:code} due to lack of space.
}

\begin{toappendix}
\ifthenelse{\boolean{useInline}}{}{
\newpage~
\newpage~
\newpage~
}
\end{toappendix}

\begin{lemmarep}\label{lem:sepa}
Given a configuration $C$ and a pattern $\hat{P}$, such that $|Q|<|\hat{P}|$, then Algorithm  {\spf}, by means of Procedure {\sepa}, requires at most one epoch in order to reach a configuration $C'$ with $|Q'|=|\hat{P}|$.
\end{lemmarep}

 
\begin{proof}
    At Line~\ref{UPFseparate}, Algorithm  {\spf} calls Procedure {\sepa} as long as $|Q|
    <|\hat{P}|$. The activated robot $r$ located at $q$, moves along its own $x$-axis of one 
    unit distance if this path is empty, otherwise it computes the position of its closest robot 
    position 
    $q''$ along its own $x$-axis, and moves toward the midpoint $z$ of $\overline{qq''}$.

    Regarding the position $q$ initially occupied by $r$, there can be two possibilities: 
    $(i)$ $q$ was occupied only by $r$; $(ii)$ $q$ was a multiplicity containing at least one more robot other than $r$.
    In the former case, after $r$ has performed its movement, the total number of robot 
    positions remains the same; in the latter, after a movement performed 
    by $r$, the number of robot positions increases by one.

    Since after each robot's movement the cardinality of $Q$ doesn't change or 
    increases by one, in order to reach a configuration $C'$ with $|Q'|=|\hat{P}|$,
    it is sufficient to activate $|\hat{P}|-|Q|$ different robots belonging to a multiplicity in 
    $C$.

    Therefore, in at most one epoch, i.e., after the activation of all the robots, we are ensured 
    to reach a configuration $C'$ where
    the number of positions occupied by the robots  $|Q'|=|\hat{P}|$.
 \end{proof}
 
\begin{lemmarep}\label{lem:last}
Let $C$ be a configuration and $\hat{P}$ a pattern such that $C$ belongs to the \emph{Finalization} stage.
Algorithm  {\spf}, by means of Procedure {\last}, finalizes the configuration in at most
$\lceil\frac \rho \delta\rceil$ epochs.
\end{lemmarep}

\begin{proof}
The \emph{Finalization} phase occurs when $Q\in {\cal B}(\hat{P})$, i.e., the pattern is formed, or when 
$(\hat{P}\setminus Q = \hat{p}_k$ $\vee$ $\hat{P}\setminus Q = \emptyset)$ $\wedge$ $(Q \setminus \hat{P} = (Q \cap \overline{O\hat{p_k}})\setminus\{\hat{p_k}\})$.
In all such cases, $|Q|\ge |\hat{P}|$ holds and Algorithm  {\spf} invokes Procedure {\last}.

If $Q\in {\cal B}(\hat{P})$, Procedure {\last} finds the correct overlapping between $Q$ and $\hat{P}$ by means of an exhaustive search at Line~\ref{last:T}.
Hence, at Line~\ref{last:F}, the returned target is the current position of the active robot, i.e., no robot moves henceforth.

In all the other cases, if the activated robot $q$ is located at $\overline{O\hat{p}_k}$, it moves toward $\hat{p}_k$, see Line~\ref{last:move}.
In each epoch, all the movements performed by the activated robots maintain the properties to belong to the \emph{Finalization} stage since such movements are along $\overline{O\hat{p}_k}$. 
Once all the moved robots reach $\hat{p}_k$, we have that $Q\in {\cal B}(\hat{P})$.

Since the robots perform non-rigid movements, i.e., they move at least $\delta$ distance, it follows that in at most $\lceil\frac {|\overline{O\hat{p}_k}|} \delta\rceil\leq \lceil \frac \rho \delta\rceil$ epochs the pattern is formed.
\end{proof}

\begin{lemmarep}\label{lem:lead}
Let $C$ be a configuration in stage \emph{Leader Configuration}. Algorithm  {\spf}, by means of Procedures {\overl} and {\lead}, requires at most $\lceil \frac \rho \delta \rceil +1$ epochs to reach a configuration $C'$ that is either in the \emph{Partial Pattern Formation} or in the \emph{Finalization} stages.    
\end{lemmarep}

\begin{proof}
Configuration $C$ is in stage \emph{Leader Configuration} whenever $|Q|\ge k$ and Procedure {\last} 
returns $null$, i.e., Algorithm {\spf} executes the Else branch of Line~\ref{spf:else}. The intent of 
the stage is to move robots so as to ensure a unique overlap between $Q$ and $\hat{P}$. In order to 
check whether such a unique overlap exists, Procedure {\overl} is invoked. 
This verifies whether $Q$ either $(i)$ admits a leader angular sequence or $(ii)$ not. In either case, 
since by hypothesis $C$ belongs to the \emph{Leader Configuration} stage, then Procedure {\lead} is invoked.

When $(i)$ holds, it means that $Q$ admits a leader angular sequence but $\Gamma$ does not. In this 
case, Procedure {\overl} exploits the cyclic order $\psi(\hat{P})$ determined by $\hat{\mu}$ and the 
leader angular sequence of $Q$ in order to obtain $\Gamma$. 
In particular, the overlap of $Q$ with $\hat{P}$, creating the joint configuration $\map$,
is done by overlapping the following: 
for $Q$, a robot position $q_j$ such that it is the second robot position determining $\theta_1$ in 
$\lambda(Q)$;
for $\hat{P}$, the first pattern point $\hat{p}_i$ on the $SEC(\hat{P})$ in the cyclic order $\psi(\hat{P})$ determined by $\hat{\mu}$. 
By definition, $\lambda(Q)$ and $\hat{\mu}$ are uniquely identifiable by all the robots. 
In case of $\hat{P}$ being symmetric, any 
selected point $\hat{p}_i$ would be equivalent with respect to positions occupied.
Moreover, the robots 
also take into consideration the clockwise direction induced by $\lambda(Q)$ and they compute the 
overlap in such a way that the order of pattern points induced by $\hat{\mu}$ follows such 
clockwise direction. Therefore, each robot computes the joint configuration 
$\map$ in the exact same way.

However, Procedure {\lead} is invoked because the resulting joint configuration 
does not admit a leader angular sequence. This means that either $\Gamma$ does not admit a unique 
smallest angle, or that the smallest angle induced by $\Gamma$ is not defined by two robot positions. 
The execution of Procedure {\lead} either makes a robot move toward $O$, or from $O$ toward a point $u'$, that guarantees a leader 
angular sequence for $\Gamma$. Point $u'$ is chosen at distance $\rho/2$ from $O$, along a radius that 
forms an angle with a robot position on the $SEC$ which measures $1/3$ of the smallest angle present in 
$\theta(S')$, with $S'$ induced by $\map$, hence creating one unique smallest angle. This induces a leader angular sequence for 
$\Gamma$. 
The existence of a robot that can move toward $O$ is guaranteed by the fact that $k\ge 5$.
The number of epochs required for the whole process is at most $\lceil\frac 3 2 \frac \rho \delta\rceil$, where 
$\frac 3 2 \rho$ is the length of the longest path that a robot may trace to reach $O$ and then 
$u'$. It is worth noting that during the movement toward $O$, angles do not change and the configuration 
maintains the described properties. In fact, any activated moving robot, is selected 
in such a way that it has a free path toward $O$, it does not break the $SEC(Q)$ and does not 
break the smallest unique angle $\theta_1$.

When $(ii)$ holds, it means that $Q$ does not admit a leader angular sequence and hence Procedure 
{\overl} returns $null$ and Procedure {\lead} is invoked. Again a robot at the center $O$ is needed to 
create the smallest unique angle in the configuration. If no robot is at the center, Procedure {\lead} 
selects any robot not responsible for the $SEC(Q)$ with a free path toward $O$, to occupy $O$. Once 
the robot at $O$ is activated, it computes the smallest angle $\xi$ in $\theta(S')$ induced by $Q$ and 
it selects as its destination the point $u'$ computed as previously described. In this case 
as well, the overall cost of Procedure {\lead} is at most $\lceil\frac 3 2 \frac \rho \delta\rceil$ epochs.

In both cases $(i)$ and $(ii)$, as soon as the robot at $O$ moves toward $u'$, the smallest unique angle is created and the configuration admits the required leader angular sequence, regardless of the 
robot reaching $u'$. 
Therefore, after reaching $O$ in at most $\lceil \frac \rho\delta \rceil$ epochs, just one more epoch is 
necessary to reach
a configuration $C'$ either in \emph{Partial Pattern Formation} or in \emph{Finalization}.
\end{proof}

\begin{lemmarep}\label{lem:walker}

Let $C$ be a configuration in stage \emph{Partial Pattern Formation}. Algorithm  {\spf}, by means of Procedures {\occ}, requires at most $2n\lceil \frac \rho \delta \rceil$ epochs to reach a configuration $C'$ that is in the \emph{Finalization} stage.    
\end{lemmarep}

\begin{proof}
A configuration $C$ is in the \emph{Partial Pattern Formation} stage whenever Procedure {\last} returns $null$ and there is a joint 
configuration $\map$ with a leader angular sequence. The goal of this stage is to make the robots move to occupy the pattern points. 
The robot $r$
designated to move in each epoch, called walker, is selected in such a way that: $r$ never crosses another robot position if there are still empty pattern points, i.e., it does not 
create or becomes part of a multiplicity; $r$ does not create a new smallest angle possibly changing the leader angular sequence, since it 
only moves following a radial detour, i.e., first toward the 
center $O$ if its target pattern point $p_l$ is located on a different radius, otherwise directly toward $p_l$; 
$r$ does not position itself 
on the path of a future walker robot, i.e., if the radius $\rho_l$ where the target pattern point
$p_l$ is located contains robot positions, $r$ is the highest priority 
free robot in $\rho_l$ that has a free path toward $p_l$. Those conditions to select the walker robot, ensure that each movement of 
$r$ makes it closer to its target pattern point $p_l$. In at most $\lceil \frac \rho \delta \rceil$ epochs $r$ reaches the center $O$ and, 
again in at most $\lceil \frac \rho \delta \rceil$ epochs, it reaches $p_l$. Therefore, in at most $2\lceil \frac \rho \delta \rceil$ epochs each 
robot moves and reaches its target pattern point. Since only one robot is selected as the walker in each epoch, after at most $2n\lceil \frac \rho \delta \rceil$ epochs, a configuration $C'$ in the \emph{Finalization} stage is reached. In fact, $C'$ is reached as soon as the last moved robot reaches $O$. Note that, the existence of a robot that can move is guaranteed by the fact that $k\ge 5$, i.e., $3$ might be stuck to maintain the $SEC$ and $1$ to maintain $\theta_1$.

Moreover, it is important to consider the two orders of priority in $\map$: the order of priority of the robots and the order of priority of the 
pattern points. The first one is given by the radiangular distance that separates each robot from $q_j$, i.e., the robot positioned on the 
$SEC(Q)$ adjacent to the smallest angle $\theta_1$. With such an ordering, the robot(s) fixing $\theta_1$ with $q_j$ is (resp., are) the 
last one (resp., ones) to become the walker. 
Let us consider the case where the walker robot $r$ is located on such a position. Then, either $(i)$ $r$ is not located at a 
multiplicity or $(ii)$ $r$ is located at a multiplicity. In the first case $(i)$, once $r$ reaches $O$, the smallest angle $\theta_1$ disappears 
and the configuration falls in the \emph{Finalization} stage. This happens because once the least priority robot moves, all the higher 
priority robots have already occupied their target pattern point and it only 
remains to occupy the last empty pattern point if $|Q|=|\hat{P}|$, or to position $r$
on a pattern point already occupied if $|Q|>|\hat{P}|$. 

In the second case $(ii)$, the walker robot $r$ moves and reaches $O$ and it does so 
without destroying $\theta_1$ since $r$ was located on a multiplicity point and such a point still maintains $\theta_1$ with $q_j$. After 
reaching $O$, $r$ moves toward $p_l$. 
Once all the robots except the last one moved out of the multiplicity point to reach their target pattern point, case $(i)$ happens and $r$ is free to move first toward $O$, and then toward $p_l$.

Finally, recall that Procedure {\occ} is executed only if $\map$ admits a leader angular 
sequence and, as we have just stated, this is maintained for the entire 
execution of the Procedure. This implies that for the entire \emph{Partial Pattern Formation} stage, the order of priority of the robots and the order 
of priority of the pattern points are uniquely identified by the activated robots in each epoch.
\end{proof}

\begin{theoremrep}\label{thm:upf}
    For every pattern $\hat{P}$ with $|\hat{P}|\ge 5$, Algorithm  {\spf} correctly solves {{\tt UPF}$^*$}  in at most $2(n+1) \lceil\frac \rho \delta\rceil +2$ epochs under \seq.
\end{theoremrep}

\begin{proof}
Let $C$ be the initial configuration. As a preliminary step, Algorithm  {\spf} takes care of $Q$, ensuring that $|Q|\ge |\hat{P}|$. By Lemma~\ref{lem:sepa} we know that this is verified after at most one epoch.

Subsequently, the algorithm checks by means of Procedure {\last} whether the configuration is already in 
the \emph{Finalization} stage or not. If this is the case, by means of Lemma~\ref{lem:last} we
know that the algorithm  
finalizes the configuration in at most $\lceil\frac \rho \delta\rceil$ epochs.

If, instead, Procedure {\last} returns $null$, the 
configuration falls into $(i)$ the \emph{Leader Configuration} stage or into $(ii)$ the \emph{Partial Pattern Formation} stage. 

In case $(i)$, according to Lemma~\ref{lem:lead}, {\spf}, by means of Procedure {\lead}, requires at most $\lceil\frac \rho \delta\rceil +1$
epochs to reach a configuration in the \emph{Partial Pattern Formation} stage as in case $(ii)$ or in the \emph{Finalization} stage. 

In case $(ii)$, according to Lemma~\ref{lem:walker}, by means of Procedure {\occ}, {\spf} requires at most 
$2n\lceil\frac \rho \delta\rceil$ epochs to reach a configuration in the \emph{Finalization} stage.

In either cases, once the \emph{Finalization} stage is reached, again according to Lemma~\ref{lem:last}, in 
at most $\lceil\frac \rho \delta\rceil$ epochs, the configuration is finalized.

Overall, Algorithm {\spf} requires at most 
$2(n+1) \lceil\frac \rho \delta\rceil +2$ epochs, to solve {{\tt UPF}$^*$} under \seq.
 \end{proof}

\section{Solving {\tt UPF}$^*$ :\ \  $1<|\hat{P}|<5$}
\label{sec:P5}
%
In this section, we study the \texttt{\Uni\ Pattern} \texttt{Formation} problem under the $\mathcal{SEQ}$ scheduler in the case $1<|\hat{P}|<5$,
and prove that it is solvable even if the robots have no additional capabilities.
%

\subsection{Solution Strategy and Algorithm}
\label{sec:smpf}
The strategy to solve the case $1< |\hat{P}| < 5$ is called {\smpf}, and similarly to {\spf} follows four stages: \emph{Initialization}, {\em Unique Maximum}, {\em Equalization}, and \emph{Finalization}. An activated robot $r$ does the following:
\begin{enumerate}
    \item If $|Q| < |\hat{P}|$, $r$ executes Procedure {\sepa} to make $|Q|\geq |\hat{P}|$ ({\em Initialization} stage);
    \item If $Q$ does not admit a unique pair of points defining the maximum distance among any pair of points in $Q$, then the first activated robot $r$ located at one of the points that define the current maximum distance, makes a move so as to enlarge its maximum distance ({\em Unique Maximum} stage);
    \item Once $Q$ admits a unique maximum distance defined by the pair $\{q_1,q_2\}$, then:
    \\
    - If $|Q|>|\hat{P}|$ then any robot not in $\{q_1,q_2\}$  
    moves toward the closest point among $q_1$ and $q_2$ ({\em Equalization} stage);
    \\
    - Else (i.e., $|Q|=|\hat{P}$), 
    $r$ suitably overlaps $\hat{P}$ with $Q$ so as to make the pair $\{q_1,q_2\}$ coincident with one of the pairs $\{\hat{p}_i,\hat{p}_j\}$ of $\hat{P}$ that define the maximum distance in $\hat{P}$.
    The {\em Finalization} stage can then start to finalize the formation of the pattern.
\end{enumerate}

In the {\em Finalization} stage, when $|Q|=|\hat{P}|$ and $Q$ admits a unique  maximum distance defined by the pair $\{q_1,q_2\}$, the overlap is computed as follows. Without loss of 
generality, let $\hat{p}_1$ and $\hat{p}_2$ be two pattern points (not necessarily unique) that define the maximum distance in $\hat{P}$. The activated robot 
$r$ overlaps the pattern by scaling it in such a way that $\overline{q_1q_2}$ and $ \overline{\hat{p}_1\hat{p}_2}$ coincide.
The pattern is considered at this scale. Now, if the activated robot $r$ is located at $q_1$ or $q_2$, it does not move. 
Consider $r$ not located at $q_1$ or $q_2$. 
There are multiple ways to overlap $\hat{P}$ with $Q$.
Since there are at most $4$ points in the pattern, there can be at most three pairs in $\hat{P}$ that can admit the maximum distance (three vertices of an equilateral triangle). Hence, the total number of distinct overlaps with rotation and reflection is at most $12$. 
The activated robot computes the pattern with smallest distance deviation, i.e., the sum of distances between pattern points and robot positions. 
If there are two overlaps of the pattern that have the same smallest distance deviation, then $r$ can consider one of them indiscriminately.

We define the distance deviation between two sets of points in $Q$ and $P$ (the corresponding overlap) as follows. Let $(q_1, q_2)$ be the unique maximum distance pair, and let $(p_1, p_2)$ be the corresponding maximum distance pair in the pattern. The distance deviation is given by:
\begin{enumerate}
    \item $|\overline{p_3q_3}|$ if $|Q| = |\hat{P}|=3$;
    \item $\min(|\overline{p_3q_3}| + |\overline{p_4q_4}|, |\overline{p_4q_3}|+ |\overline{p_3q_4}|)$ if $|Q| = |\hat{P}|=4$.
\end{enumerate}

Now, for each unoccupied pattern point, $r$ determines the closest robot position, and moves toward the unoccupied pattern point from which it is the closest. Notice that, it does not move toward the empty pattern point that it is closest to, since there may be another robot that is closer to that pattern point.

\begin{algorithm}[h!t]
\KwIn{$\hat{P}$, $Q$}
\KwOut{Destination of the robot}
$q \gets$ activated robot position\;
destination  $\gets q$\;
\eIf{$|Q| < |\hat{P}|$}{
    destination  $\gets $ {\sepa}()\; \label{UPF'separate}
}{
\eIf{$Q$ does not admit one unique maximum distance pair\label{UPF'nomax}}
{ 
\If{$q$ is one of the points among all possible pairs defining the maximum distance in $Q$\label{UPF'nomaxrobo}} 
{let $q'$ be one of the farthest points from $q$ and $\ell$ be the line passing through $q$ and $q'$\;\label{UPF'nomaxcond1}
let $q''$ be the point on $\ell$ at one unit distance from $q$, farthest from $q'$ with respect to $q$\;\label{UPF'nomaxcond2}
destination  $\gets q''$\;}\label{UPF'nomaxdest}
}{
$\{q_1, q_2\} \gets$ unique maximum distance pair of $Q$\;\label{UPF'max1}
\If{$|Q|>|\hat{P}|$\label{UPF'maxgreater}}
{
\If{$q\not \in \{q_1, q_2\}$\label{UPF'maxdef}
}{
Let $q'\in \{q_1,q_2\}$ be the closest point to $q$\;\label{UPF'maxcomp}
\If{$\overline{qq'}$ does not contain other robot positions\label{UPF'maxdestif}}{
destination  $\gets q'$\;\label{UPF'maxdest}}
}}
\Else{
    \If{$q\not \in \{q_1, q_2\}$ \label{UPF'scalerobo}}{
    \For{any pair $\{\hat{p}_1, \hat{p}_2\}$ defining the maximum distance of $\hat{P}$\label{UPF'scalepatt}}
    {Scale $\hat{P}$ such that $\overline{q_1 q_2}$ and $  \overline{\hat{p}_1\hat{p}_2}$ coincide\;\label{UPF'scale}
    \For{any rotation and reflection of $\hat{P}$ with respect to $\overline{q_1 q_2}$\label{UPF'over}} {determine an 
    overlap pattern with smallest distance deviation\;
    $p_i \gets $ the unoccupied pattern point such that $q$ is the closest robot position to $p_i$ not on a pattern point\; \label{UPF'overunocc}
\If{$\overline{qp_i}$ does not contain other robot positions}{
    destination $\gets p_i$\;\label{UPF'overdest}}
}}}}}}
\caption{{\smpf}()}\label{algo:upf2}
\end{algorithm}



\subsection{Correctness and Complexity of {\smpf}}\label{sec:corrSMPF}
We now show the correctness of the proposed Algorithm {\smpf} and discuss its complexity. 
First of all, we observe that if the input configuration admits a number of robot positions smaller than $|\hat{P}|$, then the algorithm at Line~\ref{UPF'separate} applies Procedure {\sepa} so as to obtain $|Q|=|\hat{P}|$.  By Lemma~\ref{lem:sepa}, we have that this preliminary step costs at most one epoch.

\begin{lemmarep}\label{lem:sepasmall}
    Given a configuration $C$ with $|Q| \geq |\hat{P}|$, Algorithm {\smpf}, in at most one epoch, results in a configuration $C'$ with $Q'$ admitting one unique maximum distance pair.
\end{lemmarep}
\begin{proof}
    Considering a configuration $C$ with $|Q| \geq |\hat{P}$, if Line~\ref{UPF'nomax} is true, 
    once a robot $r$ located on $q$ and defining a maximum distance in $Q$ is activated, by means of Algorithm {\smpf}, at Line~\ref{UPF'nomaxrobo} $r$ selects as its destination the point $q''$ computed as described at Lines~\ref{UPF'nomaxcond1}--\ref{UPF'nomaxcond2}. In particular, among all its most distant points in $Q$, $r$ selects one $q'$ and let  $\ell$ be the line passing through $q$ and $q'$. Then $q''$ is the point on $\ell$ at one unit distance from $q$, farthest from $q'$ with respect to $q$.
 Since $r$ was defining the previous maximum distance in $Q$, computing as its destination $q''$ means that $|\overline{q'q''}| > |\overline{q'q}|$. 
    Moreover, by means of the triangular inequality, we know that the increment of the distance 
    from $q'$, is greater than the increment of the distance from any other point. Hence, $|\overline{q'q''}|$ becomes the unique maximum distance in the reached configuration $C'$. 
\end{proof}

Another property that will be useful for our discussion, outcoming from the standard triangular inequality, concerns three points in the Euclidean plane:
\begin{property}\label{pro:triangle}
    Let $A$, $B$, and $C$ be three points in $\mathbb{R}^2$ such that $|\overline{AB}| > |\overline{AC}|$ and $|\overline{AB}| > |\overline{BC}|$. For any point $A'$ on $\overline{AC}$, $|\overline{A'B}| < |\overline{AB}|$.
\end{property}

\begin{lemmarep}\label{lem:qequalp}
    Given a configuration $C$ with a unique maximum distance pair $\{q_1, q_2\}$ and a pattern $\hat{P}$ such that 
    $|Q|>|\hat{P}|$, Algorithm {\smpf} achieves a configuration $C'$ with $|Q'|=|\hat{P}|$ in 
    at most $(n-2)\lceil\frac{|\overline{q_1q_2}|}{\delta}\rceil$ epochs.
\end{lemmarep}
\begin{proof}
    Given the assumptions, Algorithm {\smpf} executes the Lines from~\ref{UPF'max1} to~\ref{UPF'maxdest}. In particular, once a robot $r$ located at $q\not \in \{q_1, q_2\}$ is 
    activated, it selects as its destination the closest point $q'$ between $q_1$ and $q_2$.
    Thanks to Property~\ref{pro:triangle} and to the hypothesis, such a movement performed by $r$ cannot create a new maximum distance between any pair of points in $Q$. Moreover, robot $r$ 
    moves only if it has a free path toward $q'$ (see Line~\ref{UPF'maxdestif}).
    As long as there are robot positions such that $|Q|>|\hat{P}|$, Algorithm {\smpf}  
    makes the robots not located in $\{q_1, q_2\}$ move in each epoch. 
    Since there can be at most $n-2$ of such robots, all aligned toward a same target with $|\hat{P}|=2$, after at most $(n-2)\lceil\frac{|\overline{q_1q_2}|}{\delta}\rceil$ epochs, a configuration 
    $C'$ with $|Q'|=|\hat{P}|$ is achieved.
\end{proof}

\begin{lemmarep}\label{lem:invariantmaxdistance}
    Given a configuration $C$ with a unique maximum distance pair $\{q_1,q_2\}$ and a pattern $\hat{P}$ such that $|Q|=|\hat{P}|$, Algorithm {\smpf},
    in at most $2\lceil\frac{|\overline{q_1q_2}|}{\delta}\rceil$ epochs, either finalizes the pattern formation or brings to a configuration $C'$ with $|Q'|>|\hat{P}|$.
\end{lemmarep}
\begin{proof}
    According to the given assumptions, once a robot $r$ located at $q \not\in \{q_1,q_2\}$ is activated (see Line~\ref{UPF'scalerobo}), it scales the pattern $\hat{P}$ in such a way that, the maximum distance defined 
    by any pair of pattern points, has the same length of $|\overline{q_1q_2}|$ (see Lines~\ref{UPF'scalepatt} and~\ref{UPF'scale}). After that, $r$ executes Line~\ref{UPF'over} 
    computing all the possible transformations of $\hat{P}$ with respect to $\overline{q_1q_2}$. The transformation selected to overlap $Q$ with $\hat{P}$ is the one with the smallest distance deviation. Robot $r$ then computes the point $p_i$ as 
     the unoccupied pattern point such that $q$ is the closest robot position to $p_i$ not on a pattern point (see Line~\ref{UPF'overunocc}) and selects $p_i$ as its destination in Line~\ref{UPF'overdest}. 
    As long as there are unoccupied pattern points, Lines~\ref{UPF'overunocc}--\ref{UPF'overdest} 
    are executed by the activated robot $r$ to finalize the pattern. 
    Note that, there can be at most two robot positions not coincident with pattern points. Such robot positions can be aligned with a target. However, because of the maximum distance defined by $\overline{q_1 q_2}$, no other robot position nor pattern point can be aligned with $\overline{q_1 q_2}$ and outside such a segment.  
    Therefore, in at most  two epochs, Algorithm {\smpf} makes the robots not located at $\{q_1, q_2\}$ move. 

Furthermore, it is possible that after one movement a robot leaves a multiplicity, hence increasing the number of robot positions, leading to a configuration $C'$ with $|Q'|>|\hat{P}|$. 

If no multiplicities are involved, after at most $2\lceil\frac{|\overline{q_1q_2}|}{\delta}\rceil$ epochs, all the pattern points are occupied and the pattern formation is finalized.
\end{proof}

\begin{theoremrep}
    Given a configuration $C$ and a pattern $\hat{P}$ such that $|\hat{P}| \leq 4$, Algorithm {\smpf} correctly solves {\tt UPF}$^*$ in at most 
    $2(n-2)\lceil\frac{|\overline{q_1q_2}|}{\delta}\rceil+2$ epochs, with $\{q_1, q_2\}$ being the unique maximum distance pair, possibly created during the execution.
\end{theoremrep}

\begin{proof}
Algorithm {\smpf} may require  Procedure {\sepa} for one epoch in order to  obtain $|Q|=|\hat{P}|$. At most another epoch is consumed to achieve one unique maximum distance pair $\{q_1, q_2\}$. As long as $|Q|>|\hat{P}|$ or there are multiplicities not on pattern points, then, by Lemma~\ref{lem:qequalp}, at most $(n-2)\lceil\frac{|\overline{q_1q_2}|}{\delta}\rceil$ epochs to reach a configuration with a number of robot positions $|Q|=|\hat{P}|$ are required. Actually, the algorithm may alternate periods where $|Q|=|\hat{P}|$ with periods where $|Q|>|\hat{P}|$, still depending on whether there are multiplicities or not. Overall, by combining Lemmata~\ref{lem:qequalp} and~\ref{lem:invariantmaxdistance}, the algorithm requires at most $2(n-2)\lceil\frac{|\overline{q_1q_2}|}{\delta}\rceil$ to finalize the pattern. Hence, by summing up among all the possible stages of the algorithm, the claim holds.
\end{proof}

\section{{\tt Gathering} under Sequential Schedulers}\label{sec:gathering}
In the previous section, we have seen how to solve {\tt UPF}$^*$ (i.e., for $|\hat{P}|>1$) under \seq. We now consider 
the case $|\hat{P}|= 1$, that is, {\tt Gathering}. Recall that this problem is called   \texttt{Rendezvous} when there are only two robots (i.e., $|{\cal R}|=2$). 
In the literature, {\tt Gathering} has been extensively studied. However, one of the main assumptions that has been considered was to start from initial configurations without multiplicities. 
Despite such an assumption, still some known results can be easily extended to the more general {\tt Gathering} studied here, where initial configurations can admit multiplicities.

For instance, in {\fsync}, {\tt Gathering} (and hence \texttt{Rendezvous} as well) can be easily solved by the {\tt Go\_to\_the\_center\_of\_SEC} algorithm (see, e.g., \cite{CohenP05}) that makes robots move toward the center of the current $SEC$. 

Concerning {\ssy}, it is known that {\tt Rendezvous}, and hence {\tt Gathering}, are unsolvable (see, e.g.,~\cite{YamasS10}).

In {\seq} (actually in any sequential scheduler), {\tt Rendezvous} can be easily solved by making the active robot move toward the position occupied by the other robot.
Although the moving robot may not reach the destination, due to the non-rigid movements, still the distance between the two robots decreases of at least $\delta$. Hence, as soon as the distance becomes smaller than or equal to $\delta$, we are ensured that {\tt Rendezvous} is accomplished.
Without  any additional robot capability, we can prove that this result however does not extend to {\tt Gathering}.
 In fact, we show that, under a sequential scheduler,  \texttt{Gathering} of more than two robots cannot be solved in general.

\begin{theoremrep} \label{ImpossibilityGathering}  
\texttt{Gathering} is unsolvable under a sequential scheduler.
\end{theoremrep}

\begin{proof}
  By contradiction, let
     $\mathcal{A}$ be an algorithm that solves \texttt{Gathering} under a sequential scheduler ${\cal S}\in$ {\sc Sequential}.

      Consider a configuration $C(t)$ with $|{\cal R}|> 2$ and let $Q(t)$ be the corresponding set of distinct points.  
   A  robot $r$ activated at time $t$  decides  its destination  based on the observed set of points $Q(t)$;  as a result, there are only  three possible outcomes: 
  ($Act_1$) not moving;
 ($Act_2$) moving to a point occupied by another robot;
     ($Act_3$) moving to a point not occupied by another robot. 
  Since the robots have no multiplicity detection capabilities, this decision  is taken by a robot without knowing  whether or  not there are other robots located at its own position.
 We shall examine now the execution of  $\mathcal{A}$ starting from $Q(t)$.
 
  Consider first  the case where 
$|Q(t)| > 2 $. Because of the assumed correctness of ${\cal A}$,  and since at most one robot is moving in a round, there must exist a time $t'\geq t$ where  $|Q(t')| = 2$.

 Consider then the case  $|Q(t)| = 2 $. Let $A$ and $B$ be the two points of $Q(t)$;  since  $|Q(t)|= 2 < |{\cal R}|$, the  configuration must have at least one multiplicity, say $A$. 
Consider  the scenario where at any time $t'\geq t$, as long as  $|Q(t')| = 2 $, all robots co-located at $A$ (resp. $B$)  have the same coordinate system, which is
        reflexive to that  of those co-located at $B$ (resp. $A$). In this scenario, which we shall call  {\em mirror}, any activated robot will make the same decision as long as  $|Q(t')| = 2 $.
         
  Observe that, regardless of where the activated robot $r$ resides, the decision according to $\mathcal{A}$  cannot be $Act_1$, i.e., not moving. 
  Otherwise, in the execution of  $\mathcal{A}$ in the mirror scenario, no activated robot will ever move; that is,
        $\forall t'> t,\  Q(t') = Q(t)=2$ and \texttt{Gathering} remains unsolved, contradicting the assumed correctness of  $\mathcal{A}$.

Moreover, the decision by a robot according to $\mathcal{A}$  cannot be $Act_2$, i.e., move to $B$.  
Consider in fact the execution of $\mathcal{A}$ in the mirror scenario where
 the scheduler  activates sequentially  all robots but one (say $a$) from $A$, 
 then sequentially all the robots originally in $B$,  and finally $a$. In so doing, the  reached configuration is another mirror configuration, where the positions of the robots have now switched. 
 Observe that, at any time $t'$ during this process, $Q(t')=2$.  From this moment on,  the scheduler continues to perform these switches. Thus, again $\forall t> 0,\  Q(t) = Q(0)=2$, and \texttt{Gathering} remains unsolved, contradicting the assumed correctness of $\mathcal{A}$.

It remains to analyze        
 $Act_3$, i.e., moving to a point distinct from $A$ and $B$, say $C$;    
if the activated robot located in either $A$ or $B$ is part of a  multiplicity, then $|Q(t+1)|=3$, otherwise $|Q(t+1)|=2$.  In other words,
in any execution of  $\mathcal{A}$, if $|Q(t)|=2$, then  $ |Q(t+1)|\geq 2$.

Summarizing, $\forall t\geq 0,\ |Q(t)|>1$; that is,  in any execution of  $\mathcal{A}$ the robots never gather,
contradicting the assumed correctness of  $\mathcal{A}$. 
\end{proof}

We have just seen that without any additional capability, {\tt Gathering} is unsolvable under any sequential scheduler. 
Here we consider the case where the capability of the robots is empowered with weak multiplicity detection, i.e., robots are able to distinguish whether a point is occupied by a single robot or by a multiplicity (but not the exact number of robots composing the multiplicity).
More precisely, any  robot 
$r$ activated at time $t$ detects the set $Q'(t)=\{(q_1,b_1), \dots, (q_m,b_m)\}$ where $b_i\in \{0,1\}$ indicates whether $q_i$ contains a multiplicity ($b_i=1$) or not ($b_i=0$).

Let $r$ be the robot activated at time $t$ and $\kappa$ be the number of multiplicities in $Q'(t)$; the action of $r$ dictated by Algorithm~\ref{algo:seqgather} {\sc SqGathering} follows a very simple principle, depending on $\kappa$:

\begin{enumerate} 
    \item$\kappa= 1$: If $r$ is at the multiplicity, do not move. Otherwise, move to the point with multiplicity;
    \item$\kappa > 1$: If $r$ is at a multiplicity, move to a point not occupied by any other robot. Otherwise, do not move;
    \item$\kappa = 0$: Move to the closest point occupied by another robot.
\end{enumerate}

\begin{algorithm}[h!t]
\KwIn{$Q'(t)$}
\KwOut{Destination of the robot}
$(q, b) \gets$ position of the activated robot\;
$\kappa \gets$ number of multiplicities in $Q'(t)$\;
destination $\gets q$\;
\If{$\kappa = 1$}{
    \If{$b = 0$}{
        $q_m \gets$ the multiplicity point in $Q'(t)$\;
        destination $\gets q_m$\;
    }
}
\Else{\If{$\kappa  > 1$}{
    \If{$b = 1$}{
        $q' \gets$ closest point to $q$\;
        $z \gets$ midpoint of $\overline{qq'}$\;
        destination $\gets z$\;}
    }
\Else{$q' \gets$ closest point to $q$\;
    destination $\gets q'$\;
}}
\caption{{\sc SqGathering}}\label{algo:seqgather}
\end{algorithm}


 \begin{theoremrep} 
 {\tt Gathering} is  solvable   with weak multiplicity detection under any sequential scheduler. 
\end{theoremrep}
\begin{proof}
    Algorithm {\sc SqGathering} solves {\tt Gathering} by maintaining a unique multiplicity point and moving other robots sequentially to that multiplicity point. If there is no multiplicity, then the activated robot creates a multiplicity. If there is more than one multiplicity at time $t$, then the robots move to unoccupied points to separate such that at some time $t' > t$, there is only one multiplicity. Then for all $t'' > t'$, that multiplicity is maintained.  
\end{proof}

By contrast, the addition of  
multiplicity detection is not enough under most synchronous schedulers, as shown by the following known result:

\begin{lemmarep} 
 {\tt Gathering} under \ssy\ is unsolvable with multiplicity detection.
\end{lemmarep}
\begin{proof}
    From \cite{YamasS10}, {\tt Rendezvous} is unsolvable under \ssy. Such a result can be easily extended to the case where robots are empowered with 
    weak multiplicity detection. It is sufficient to start from a configuration with two multiplicities. Similarly, the result can be extended 
    to the case where robots are empowered with the strong multiplicity detection, i.e., when they recognize how many robots compose a 
    multiplicity. In this case it is sufficient to start from a configuration with two multiplicities composed by the same number of robots.
\end{proof}

\section{Weighted Pattern Formation}\label{sec:wpf}

In this section we consider the \emph{Weighted Pattern Formation} (\texttt{WPF}) problem, in which each point of the target pattern is associated with a \emph{weight}, i.e., then number of robots necessary on that point to be considered occupied (i.e., its multiplicity). A \emph{weighted pattern} is a pair $(\hat{P},\hat{W})$, where $\hat{P}=\{\hat{p}_1,\hat{p}_2,\dots,\hat{p}_k\}$ is a pattern and $\hat{W}=\{\hat{w}_1,\hat{w}_2,\dots,\hat{w}_k\}$ is a set of positive integers, with $\hat{w}_i$ denoting the desired number of robots to be located at $\hat{p}_i$.

The robots are said to have formed the weighted pattern $(\hat{P},\hat{W})$ at time $t$ if $Q(t)\in\mathcal{B}(\hat{P})$ and the multiplicity of each occupied location matches the weight of the corresponding point in the target pattern.

Formally, \texttt{WPF} is defined by the following temporal geometric predicate:
 \begin{align*}
     {\tt WPF} &\equiv \forall   \hat{P}\subset \mathbb R^2, \hat{W}\subset \mathbb Z^+ : |\hat{P}|=|\hat{W}|>1,  \forall n = \sum_{w\in \hat{W}}  w,\   
     \\&\forall C(0)\in \mathbb R^2_n,\   \exists t\geq 0 :
     (Q(t)\in {\cal B}(\hat{P}))\  \textbf{and}\  
     \\&( \forall q\in Q(t),  m(q)=w(\iota(q)))\ \textbf{and}\     (\forall t'>t, C(t') = C(t)),
 \end{align*}
where $m(q)$ denotes the multiplicity of location $q$, $\iota(q)$ is the point of $\hat{P}$ corresponding to $q$ under the isomorphism, and $w(p)$ denotes the weight associated with the pattern point $p$.

Observe that solving \texttt{WPF} requires robots to be equipped with \emph{strong multiplicity detection}, since they must distinguish the exact number of robots occupying every location. Consequently, they also know the total number of robots,
$n=\sum_{w\in\hat{W}}w$.

\subsection{Resolution Strategy}

First we define different types of pattern points in order to make it clear which are the ones that still need to reach the required multiplicity and which are the ones where too many robots are located at.

We say that a pattern point is \emph{deficient} if the number of robots at the pattern point is smaller than the weight of the pattern point.
Similarly, the pattern point is \emph{excess} if the number of robots at the pattern point is greater than the weight.
The target pattern point at any step is the highest priority pattern point $p_i$, ranked in the same way as in the {\UPF*} strategy such that $p_i$ is deficient. 

Differently from the \emph{free} definition of a robot in the {\UPF*} strategy, here we say that a robot is \emph{weighted free} if it is not located at a pattern point or if it is located at an excess pattern point. 
The robot selected by the algorithm to move to the highest priority pattern point is called \emph{weighted walker}. Such a robot is selected applying the same conditions for the {\UPF*} walker but considering weighted free robots and deficient/excess pattern points.
The activated robot can always determine the weighted walker as long as the configuration composed of robots and pattern points is a leader configuration. 

Also the weighted walker is selected with respect to the target pattern point. This selection guarantees that the chosen robot always has an unobstructed radial detour to its destination. Moreover, if the highest priority weighted free robot can reach the target without encountering other robots, it is selected as the walker and moves at least $\delta$ distance in each activation; otherwise, priority (i.e., the selected weighted  walker) is given to the robot that is closest to the target along the corresponding radial detour. If the radius of the target pattern point $p_i$ already contains other robots in $\overline{Op_i}$, the weighted walker is chosen among the robots in $\overline{Op_i}$ so that the path to the destination remains unobstructed.

The rest of the strategy to solve {\WPF} remains the same as the one used to solve {\UPF*}.

We have the following theorem:
\begin{theorem} 
{\tt WPF} is solvable with strong multiplicity detection under any sequential scheduler. 
\end{theorem}

\section{Conclusions and Open Problems}\label{sec:conclusion}

\subsection{Summary}
 We have studied the computational  power of oblivious robots  operating  in the plane  under  {\em sequential}  schedulers.
 Focusing on  the fundamental class of pattern formation problems, we
  investigated the adversarial power of such schedulers in comparison to that of 
 the  other two important synchronous schedulers  studied in the literature: \fsync\ and \ssy.
Specifically, we considered  the resolution  of the following  
 problems:   {\tt UPF}, the  most general problem in the class;  {\tt Gathering}, or point-formation; and
${\tt UPF}^* = {\tt UPF}\ \setminus$  {\tt Gathering}.
 
We have shown that,  under any sequential scheduler,  oblivious robots can solve:
${\tt UPF^*}$, without any additional assumption; 
 {\tt Gathering} and, hence, {\tt UPF}, with the only addition of weak multiplicity detection.

\begin{figure}[t]
\centering
\resizebox{0.8\columnwidth}{!}{
    \begin{tikzpicture}
    \tikzset{every node/.style={align=center}}
    
    \draw[thick, red] (-1.5,0) ellipse (2.5 and 1) node[left=2.5cm] {$\mathcal{FSYNC}$};
    \draw[thick, blue] (1.5,0) ellipse (2.5 and 1) node[right=2.5cm] {$\mathcal{SEQ}$};
    
    \begin{scope}
        \clip (-1,0) ellipse (2 and 1);
        \clip (1,0) ellipse (2 and 1);
        \draw[thick, color={rgb,255:red,50; green,100; blue,0}] (0,0) circle (0.5);
    \end{scope}

    \node at (-3.2,0) [circle, fill=black, inner sep=1pt, label=right:{\tt Gathering}] {};
    \node at (1.7,0) [circle, fill=black, inner sep=1pt, label=right:{\tt UPF$^*$}] {};

    \node (ssy) at (0,-1.5) [circle, thick,color={rgb,255:red,50; green,100; blue,0}] {$\mathcal{SSYNC}$};
    
    \draw[-latex] (0, -1.3) -- (0,-0.5);
\end{tikzpicture}
}
\caption{Relationship between the set of problems solvable under \fsync, \seq, and \ssy, without any additional assumptions.}
\label{fig:summary}
\end{figure}

In contrast, we  proved that ${\tt UPF^*}$
is unsolvable under \fsync\ (and hence under \ssy\ and \async) even by assuming strong multiplicity detection, a leader, rigid movements, orientation.

Since it is known from the literature that {\tt Gathering} is solvable under \fsync\ but not under \ssy, then we conclude that \seq\ and \fsync\ are orthogonal in general, whereas the computational power of the robots under \seq\ is stronger than under \ssy, see Fig. \ref{fig:summary}.

\subsection{Open Problems and Extensions} 

The results of this paper indicate the strong computational power that the robots in \OB\
have under {\it any} sequential scheduler. They constitute a preliminary foundation for further examination of sequential schedulers, study of their properties, and analysis of their impact.
Indeed, a  broad spectrum of problems is opened by our study. 

The main concern of our research has been computability rather than complexity. An immediate open problem is to improve the time (i.e., epoch) complexity of our protocols,   possibly establishing tight bounds. 
Also, since the concern so far has been restricted to deterministic protocols, this leaves open the use of probabilistic means.

Our investigation has been carried out in the standard \OB\ model.
Even just within this model, our knowledge so far is limited by the standard assumption that the robots can view the positions of all other robots. In particular nothing is currently known when
the visibility of the robots operating under sequential schedulers has a limited  range (e.g., in \cite{KirkpKNPS24}) or is obstructed (e.g., in \cite{FelettiMMP25}).

Extending the investigation to  the other  established models  (${\cal L}${\sc UMI}, ${\cal F}${\sc COM}, ${\cal F}${\sc STA}) is clearly an important research direction. Following the  announcement of the results of this paper in \cite{FNPPS2025}, preliminary investigations of this topic have already started  \cite{FeFS25,FeFPPS25}.

\bibliography{references}

\begin{thebibliography}{10}
\providecommand{\url}[1]{{#1}}
\providecommand{\urlprefix}{URL }
\providecommand{\doi}[1]{\url{https://doi.org/#1}}


\bibitem{SuzukY99}
I.~Suzuki, M.~Yamashita, Distributed anonymous mobile robots: Formation of geometric patterns.
\newblock SIAM Journal on Computing \textbf{28}(4), 1347--1363 (1999)

\bibitem{CanepDIP16}
D.~Canepa, X.~D{\`e}fago, T.~Izumi, M.~{Potop-Butucaru}, \emph{Flocking with oblivious robots}, in \emph{Proc. of 18th Int.'l Symp. on Stabilization, {{Safety}}, and {{Security}} of {{Distributed Systems}} ({{SSS}})} (2016), pp. 94--108

\bibitem{FloccPS12}
P.~Flocchini, G.~Prencipe, N.~Santoro, \emph{Distributed {{Computing}} by {{Oblivious Mobile Robots}}}.
\newblock Synthesis {{Lectures}} on {{Distributed Computing Theory}} (Springer Cham, 2012)

\bibitem{IzumiSKIDWY12}
T.~Izumi, S.~Souissi, Y.~Katayama, N.~Inuzuka, X.~D{\'e}fago, K.~Wada, M.~Yamashita, The gathering problem for two oblivious robots with unreliable compasses.
\newblock SIAM Journal on Computing \textbf{41}(1), 26--46 (2012)

\bibitem{YamasS10}
M.~Yamashita, I.~Suzuki, Characterizing geometric patterns formable by oblivious anonymous mobile robots.
\newblock Theor. Comput. Sci. \textbf{411}, 2433--2453 (2010)

\bibitem{FloccPSW99}
P.~Flocchini, G.~Prencipe, N.~Santoro, P.~Widmayer, \emph{Hard tasks for weak robots}, in \emph{Proc. of 10th {{Int.'l Symp.}} Alg. {{Comput.}} ({{ISAAC}})} (1999), pp. 93--102

\bibitem{CicerDN21}
S.~Cicerone, G.~Di~Stefano, A.~Navarra, Semi-asynchronous: {A} new scheduler in distributed computing.
\newblock IEEE Access \textbf{9}, 41540--41557 (2021)

\bibitem{FloccSSW23}
P.~Flocchini, N.~Santoro, Y.~Sudo, K.~Wada, \emph{On asynchrony, memory, and communication: {{Separations}} and landscapes}, in \emph{Proc. of 27th Int.'l Conf. on {Principles of Distributed Systems} ({{OPODIS}})} (2023), pp. 28:1--28:23

\bibitem{KirkpKNPS24}
D.~Kirkpatrick, I.~Kostitsyna, A.~Navarra, G.~Prencipe, N.~Santoro, On the power of bounded asynchrony: Convergence by autonomous robots with limited visibility.
\newblock Distibuted Computing \textbf{37}, 1--30 (2024)

\bibitem{Fa2018}
N.~Fat{\`e}s, \emph{Asynchronous Cellular Automata} (Springer Berlin Heidelberg, 2018), pp. 1--21

\bibitem{DefagPT19}
X.~D{\'e}fago, M.~{Potop-Butucaru}, S.~Tixeuil, {\em Fault-{{Tolerant Mobile Robots}}}.
\newblock {Chapter 10 of} \cite{FloccPS19} pp. 234--251 (2019)

\bibitem{FreiW24}
F.~Frei, K.~Wada, \emph{{Brief announcement: Distinct gathering under round robin}}, in \emph{38th International Symposium on Distributed Computing (DISC 2024)}, vol. 319 (2024), pp. 48:1--48:8

\bibitem{NP25}
A.~Navarra, F.~Piselli, \emph{Oblivious robots under round robin: Gathering on rings}, in \emph{Frontiers of Algorithmics} (Springer Nature Singapore, 2025), pp. 166--180

\bibitem{DefagS08}
X.~D{\'e}fago, S.~Souissi, Non-uniform circle formation algorithm for oblivious mobile robots with convergence toward uniformity.
\newblock Theor. Comput. Sci. \textbf{396}(1-3), 97--112 (2008)

\bibitem{DieudLP08}
Y.~Dieudonn{\'e}, O.~{Labbani-Igbida}, F.~Petit, Circle formation of weak mobile robots.
\newblock ACM Trans. on Autonomous and Adaptive Systems \textbf{3}(4), 1--20 (2008)

\bibitem{FloccPSV17}
P.~Flocchini, G.~Prencipe, N.~Santoro, G.~Viglietta, Distributed computing by mobile robots: {{Uniform}} circle formation.
\newblock Dist. Comp. \textbf{30}(6), 413--457 (2017)

\bibitem{AgmonP06}
N.~Agmon, D.~Peleg, Fault-tolerant gathering algorithms for autonomous mobile robots.
\newblock SIAM Journal on Computing \textbf{36}(1), 56--82 (2006)

\bibitem{CieliFPS12}
M.~Cieliebak, P.~Flocchini, G.~Prencipe, N.~Santoro, Distributed computing by mobile robots: {{Gathering}}.
\newblock SIAM Journal on Computing \textbf{41}(4), 829--879 (2012)

\bibitem{Flocc19}
P.~Flocchini, {\em Gathering}.
\newblock {Chapter 4 of} \cite{FloccPS19} pp. 63--82 (2019)

\bibitem{KameiLOT11}
S.~Kamei, A.~Lamani, F.~Ooshita, S.~Tixeuil, \emph{Asynchronous mobile robot gathering from symmetric configurations without global multiplicity detection}, in \emph{Proc. of 18th {{Int.'l Colloquium}} on {{Structural Information}} and {{Communication Complexity}} ({{SIROCCO}})} (2011), pp. 150--161

\bibitem{PattaAM21}
D.~Pattanayak, J.~Augustine, P.S. Mandal, Randomized gathering of asynchronous mobile robots.
\newblock Theor. Comput. Sci. \textbf{858}, 64--80 (2021)

\bibitem{PattaMHM19}
D.~Pattanayak, K.~Mondal, R.~H., P.S. Mandal, Gathering of mobile robots with weak multiplicity detection in presence of crash-faults.
\newblock Journal of Parallel and Distributed Computing \textbf{123}, 145--155 (2019)

\bibitem{Prenc07}
G.~Prencipe, Impossibility of gathering by a set of autonomous mobile robots.
\newblock Theoretical Computer Science \textbf{384}(2-3), 222--231 (2007)

\bibitem{CicerDN19}
S.~Cicerone, G.~Di~Stefano, A.~Navarra, Embedded pattern formation by asynchronous robots without chirality.
\newblock Distributed Computing \textbf{32}(4), 291--315 (2019)

\bibitem{CicerDN19a}
S.~Cicerone, G.~Di~Stefano, A.~Navarra, Asynchronous arbitrary pattern formation: {The} effects of a rigorous approach.
\newblock Distributed Computing \textbf{32}(2), 91--132 (2019)

\bibitem{CicerSN21}
S.~Cicerone, G.~{Di Stefano}, A.~Navarra, Solving the pattern formation by mobile robots with chirality.
\newblock IEEE Access \textbf{9}, 88177--88204 (2021)

\bibitem{DPV10}
Y.~Dieudonn{\'e}, F.~Petit, V.~Villain, \emph{Leader election problem versus pattern formation problem}, in \emph{Proc. of 24th Int.'l Symp. on Distributed Computing (DISC)} (2010), pp. 267--281

\bibitem{FloccPSW08}
P.~Flocchini, G.~Prencipe, N.~Santoro, P.~Widmayer, Arbitrary pattern formation by asynchronous, anonymous, oblivious robots.
\newblock Theor. Comput. Sci. \textbf{407}(1), 412--447 (2008)

\bibitem{CohenP05}
R.~Cohen, D.~Peleg, Convergence properties of the gravitational algorithm in asynchronous robot systems.
\newblock SIAM Journal on Computing \textbf{34}(6), 1516--1528 (2005)

\bibitem{FelettiMMP25}
C.~Feletti, L.~Mambretti, C.~Mereghetti, B.~Palano, Computational power of autonomous robots: Transparency vs. opaqueness.
\newblock Theor. Comput. Sci. \textbf{1036}, 115--153 (2025)

\bibitem{FNPPS2025}
P.~Flocchini, A.~Navarra, D.~Pattanayak, F.~Piselli, N.~Santoro, \emph{Oblivious robots under sequential schedulers: Universal pattern formation}, in \emph{Proc. 32nd Int.'l Colloquium on Structural Information and Communication Complexity ({SIROCCO})} (2025), pp. 297--314

\bibitem{FeFS25}
C.~Feletti, P.~Flocchini, N.~Santoro, \emph{On the computational power of mobile robots under sequential schedulers}, in \emph{Proc. of 27th International Symposium on Stabilization, Safety, and Security of Distributed Systems (SSS)} (2025), pp. 216--232

\bibitem{FeFPPS25}
C.~Feletti, P.~Flocchini, D.~Pattanayak, G.~Prencipe, N.~Santoro, \emph{Brief announcement: Universal dancing by luminous robots under sequential schedulers}, in \emph{Proc. 39th International Symposium on Distributed Computing (DISC)} (2025), pp. 56:1--56:7

\bibitem{FloccPS19}
P.~Flocchini, G.~Prencipe, N.~Santoro~(Eds), \emph{Distributed Computing by Mobile Entities} (Springer, 2019)

\end{thebibliography}

\end{document}